\documentclass[proceedings]{JHEP35}
\usepackage{amssymb}
\usepackage{amsfonts}
\usepackage{amsmath}
\usepackage{epsfig}
\usepackage{xcolor}

\newbox\mybox

\newcommand\fverb{\setbox\mybox=\hbox\bgroup\verb}
\newcommand\fverbdo{\egroup\medskip\noindent\fbox{\unhbox\mybox}\ }
\newcommand\fverbit{\egroup\item[\fbox{\unhbox\mybox}]}
\conference{Lorentzian Toda field theories}
\abstract{We propose several different types of construction principles for new classes of Toda field theories based on root systems defined on Lorentzian lattices. 
In analogy to conformal and affine Toda theories based on root systems of semi-simple Lie algebras, also their Lorentzian extensions come about in conformal and massive variants. 
We carry out the Painlev\'{e} integrability test for the proposed theories, finding in general only one integer valued resonance corresponding to the energy-momentum tensor. Thus most
of the Lorentzian Toda field theories are not integrable, as the remaining resonances, that grade the spins of the W-algebras in the semisimple cases, are either non integer or complex valued. 
We analyse in detail the classical mass spectra of several massive variants. Lorentzian Toda field theories may be viewed as perturbed versions of integrable theories equipped with an
algebraic framework.}

\title{Lorentzian Toda field theories}
\author{Andreas Fring and Samuel Whittington \\
Department of Mathematics, City University London,\\
Northampton Square, London EC1V 0HB, UK\\
E-mail: a.fring@city.ac.uk,samuel.whittington@city.ac.uk}

\input{tcilatex}
\begin{document}

\section{Introduction}

Toda theories are one of the best studied and understood classical and
quantum integrable systems. The integrability of their classical discrete
lattice versions \cite{toda1975} is known for a long time and has been
established by the construction of explicit Lax pairs \cite{bogoyavlensky}
as well as the application of the Painlev\'{e} integrability test \cite%
{yoshida1983}. Their continuous field theoretical versions are scalar field
theories defined by Lagrangians of the general form%
\begin{equation}
\mathcal{L}_{_{\mathbf{g}_{-n}}}=\frac{1}{2}\partial _{\mu }\phi \cdot
\partial ^{\mu }\phi -\frac{g}{\beta ^{2}}\sum\limits_{i=-n}^{r}e^{\beta
\alpha _{i}\cdot \phi },  \label{1}
\end{equation}%
with coupling constants $g,\beta \in \mathbb{R}$ or $\beta \in i\mathbb{R}$.
The $r+n+1$ vectors $\alpha _{i}$ of dimension $\ell +2m$ are taken to be
roots on a lattice associated to some Lie algebras and the scalar field $%
\phi (x,t)$ has in general $\ell +2m$ components, i.e. $\phi ^{a}(x,t)$ with 
$a=1,\ldots ,\ell +2m$. Folded versions may also been constructed in which
some field components are identified in very specific ways, see e.g. \cite%
{fring2005affine}.

Many versions of the Lagrangians in (\ref{1}) have been well studied. For
instance, for $\ell =r$, $m=0$, $n=-1$ with $\alpha _{i}$ taken to be the
simple roots of a semisimple Lie algebra $\mathbf{g}$, the Lagrangians\ $%
\mathcal{L}_{_{\mathbf{g}_{1}}}$corresponds to the well known description of
conformal Toda field theory, see e.g. \cite{curt1982conf,mansfield1983light}%
. When $\ell =r$, $m=0$, $n=0$, with $\alpha _{0}$ taken to be the negative
of the highest root, the Lagrangians $\mathcal{L}_{_{\mathbf{g}_{0}}}$%
corresponds to affine Toda field theory with $r$ massive scalar fields, see
for instance e.g. \cite{BCDS,Mass2}. Similarly as for their discrete
counterparts, the classical integrability of these continuous systems has
been established by the explicit construction of Lax pairs or zero curvature
expressions \cite{toda1,OP4,wilson81,Integ}. Since the classical equations
of motion are nonlinear integrable equations, they possess solutions of with
very rich solitonic structures \cite{nt,Holl,aratyn93,David,fring1994vertex}%
. One of the most remarkable properties of these systems is the fact that
the quantum scattering matrices for affine Toda field theories have been
constructed to all orders in perturbation theory by using what is referred
to as the bootstrap approach \cite{toda2,zamo1989,CM1,CM2,BCDS,dorey91,FO}.
For theories based on simply laced Lie algebras this has been possible due
to the fact that the classical mass ratios \cite{Mass2} are preserved to all
orders in perturbation theory \cite{toda2,zamo1989,CM1,CM2,DDV}. For
non-simply laced theories the masses renormalize with different factors,
but, nonetheless, closed exact expressions for the scattering matrices were
still found \cite{q1,q2} by exploiting properties of q-deformed Coxeter
elements. Once again the root space provides the general framework, where in
this case the dual affine algebras correspond to the two classical limits of
very weak or very strong coupling. While the Yang-Baxter equation is
trivially satisfied by the scattering matrices describing theories with real
coupling constants $\beta \in \mathbb{R}$, it possesses non-trivial
solutions characterized by their quantum group symmetries when $\beta \in i%
\mathbb{R}$ is taken to be purely imaginary \cite{bernard1991}. The
S-matrices factorize into the so-called minimal and CDD factor, with the
former describing the scattering in the Restricted Solid-on-Solid
(RSOS)-models and the latter containing the coupling constant.

While all of the above theories are integrable and based on root systems
that lead to a positive definite or semi-definite Cartan matrix, some
attempts have been made to extend the theories in (\ref{1}) and formulate
them on root systems corresponding to hyperbolic Kac Moody algebras \cite%
{gebert1996toda}. These algebras have been fully classified \cite%
{carbone2010} and proven to be very useful in a string/M theory context \cite%
{damour200210}, with $E_{10}$ being a popular example. However, it has
turned that $E_{11}$, which is not a hyperbolic Kac Moody algebra \cite%
{nicolai2004low}, is even more useful. It belongs to the larger class of
algebras that are Lorentzian \cite%
{borcherds1988generalized,gaberdiel2002class,fring2019n}. A particular
subclass of them studied in \cite{gaberdiel2002class}, is defined in terms
of their connected Dynkin diagrams so that the deletion of at least one node
leaves a possibly disconnected set of connected Dynkin diagrams each of
which is of finite type, except for at most one affine type. In \cite%
{fring2019n} an even larger class of $n$-extended Lorentzian Kac Moody
algebras was introduced. Here the aim is to investigate the properties of
Toda field theories described by the Lagrangians in (\ref{1}) based on these
Lorentzian type of root systems.

Since the algebras discussed come along with a classification scheme based
on their root systems, the above results have proven very successfully that
the physical properties of the theories based on them can be characterized
very systematically. In has turned out that theories in the same subclass
share the same general properties. These subclasses may be defined for
instance by being simply laced or non-simply laced, semisimple or affine,
having real or purely imaginary coupling constants, being hyperbolic, etc.
It has turned out that Toda field theories based on root systems
corresponding to hyperbolic Kac Moody algebras are not integrable as they do
not pass the Painlev\'{e} test \cite{gebert1996toda}. However, similarly as
their integrable cousins, they, together with the theories of Lorentzian
type discussed here, provide a systematic framework for the study of
nonintegrable quantum field theories \cite{delfino1996non}. We take these
two aspects as our main motivation to study Lorentzian Toda field theories.

To set up these new systems we need to specify not only the limits in the
sum in (\ref{1}), the dimension of the representation space of the roots and
the choice of the highest root similarly as for the conformal and affine
cases, but in addition we also have to re-defined the inner product in the
kinetic term between the derivatives of the fields and in the potential term
between the roots and the scalar field. These new inner products will place
the theories onto Lorentzian lattice.

Our manuscript is structured as follows: In section 2 we introduce our
definition and some key properties of the Lorentzian inner products. In
addition, we introduce several matrices that are central for our analysis.
In section 3 we employ these products to set up our Lorentzian Toda field
theories. In section 4 we carry our the Painlev\'{e} integrability test for
these theories. The test can be entirely reduced to an eigenvalue problem of
what we refer to as the Painlev\'{e} matrix, that we analyse in section 5.
In section 6 we discuss in detail the classical mass spectra of Lorentzian
Toda field theories produced from different schemes. Our conclusions are
stated in section 7.

\section{Lorentzian products, the K, M, $\Lambda $, D and Painlev\'{e}
matrices}

The main difference between theories based on root systems for semisimple or
affine algebras is the re-definition of the inner product between the
derivatives of the scalar fields and between the roots and the scalar fields
in kinetic and the potential term in (\ref{1}), respectively. Following \cite%
{gaberdiel2002class,fring2019n}, we define here the following Lorentzian
inner products for two $\ell +2m$ dimensional vectors $x=(x_{1},\ldots
,x_{\ell +2m})$ and $y=(y_{1},\ldots ,y_{\ell +2m})$ as%
\begin{equation}
x\cdot y:=\sum\limits_{\beta =1}^{\ell }x_{\beta }y_{\beta
}-\sum\limits_{\beta =1}^{m}\left( x_{\ell +2\beta -1}y_{\ell +2\beta
}+x_{\ell +2\beta }y_{\ell +2\beta -1}\right) .  \label{vecmult}
\end{equation}%
We extend the definition of this product to matrix multiplication in a
natural way. For a $N\times \left( \ell +2m\right) $-matrix $A$ and a $%
\left( \ell +2m\right) \times N$-matrix $B$, we define%
\begin{equation}
\left( AB\right) _{ij}:=\sum\limits_{\beta =1}^{\ell }A_{i\beta }B_{\beta
j}-\sum\limits_{\beta =1}^{m}\left[ A_{i\left( \ell +2\beta -1\right)
}B_{\left( \ell +2\beta \right) j}+A_{i\left( \ell +2\beta \right)
}B_{\left( \ell +2\beta -1\right) j}\right] ,~\ i,j=1,\ldots ,N.
\label{matrixmult}
\end{equation}%
In particular, taking now $N=r+n+1$ we define a $(r+n+1)\times \left( \ell
+2m\right) $-matrix $M$ with rows comprised of $r+n+1$ root vectors $\alpha
_{i}=$ $(\alpha _{i}^{1},\ldots ,\alpha _{i}^{_{\ell +2m}})^{T}$ of
dimension $\ell +2m$, i.e. $M_{i\beta }:=\alpha _{i}^{\beta }$. When $%
r+n+1\leq \ell +2m$ the matrix $M$ possess a right inverse, which is
obtained by\ defining a $\left( \ell +2m\right) \times (r+n+1)$-matrix $%
\Lambda $ with columns comprised of $r+n+1$ fundamental weight vectors $%
\lambda _{i}=$ $(\lambda _{i}^{1},\ldots ,\lambda _{i}^{_{\ell +2m}})$ of
dimension $\ell +2m$, i.e. $\Lambda _{\beta i}:=\lambda _{i}^{\beta }$.
Hence with%
\begin{equation}
M_{i\beta }:=\alpha _{i}^{\beta },\qquad \Lambda _{\beta i}:=\lambda
_{i}^{\beta },~~\ \ \ ~i=1,\ldots ,r+n+1\text{; }\beta =1,\ldots ,\ell ,%
\text{ }  \label{M}
\end{equation}%
we obtain%
\begin{equation}
\left( M\Lambda \right) _{ij}=\alpha _{i}\cdot \lambda _{j}=\delta
_{ij}=\lambda _{i}\cdot \alpha _{j}=\left( \Lambda ^{T}M^{T}\right) _{ij}.
\end{equation}%
Moreover, we may employ $M$, $M^{T}$ and $\Lambda $, $\Lambda ^{T}$ to
factorize the symmetric Cartan matrix $K$ and their inverse $K^{-1}$,
respectively, as%
\begin{equation}
\left( MM^{T}\right) _{ij}=\alpha _{i}\cdot \alpha _{j}=K_{ij},\quad \text{%
and\quad }\left( \Lambda \Lambda ^{T}\right) _{ij}=\lambda _{i}\cdot \lambda
_{j}=K_{ij}^{-1}.  \label{fact}
\end{equation}%
In general the Cartan matrix is defined as $2\alpha _{i}\cdot \alpha
_{j}/\alpha _{j}^{2}$, which only in the symmetric case may be reduced to $%
\alpha _{i}\cdot \alpha _{j}$ when taking the length of the roots to be $2$.
Since below we shall also encounter roots of length $0$, we adopt here the
symmetric definition. When summing over one index of the inverse symmetric
Cartan matrix we obtain some constants 
\begin{equation}
D_{k}:=\sum_{j=-n}^{r}K_{kj}^{-1}=\rho \cdot \lambda _{k},~~~\ \ \
k=-n,\ldots ,-1,0,1,\ldots ,r,  \label{Di}
\end{equation}%
that encode information about the existence of $SO(1,2)$ and $SO(3)$
principal subalgebras and the decomposition of the root lattices with their
corresponding algebras \cite{gaberdiel2002class,fring2019n}. As stated in (%
\ref{Di}) the constants $D_{k}$ can also be computed directly from their
Lorentzian inner products of weight vectors $\lambda _{k}$ with the Weyl
vector $\rho $. Using these constants to define a diagonal matrix $D:=%
\limfunc{diag}(D_{r},\ldots ,D_{-n})$ we introduce a further matrix 
\begin{equation}
P:=2DK,  \label{P}
\end{equation}%
referred to here as the \emph{Painlev\'{e} matrix}. As we will see below the
Painlev\'{e} integrability test can be reduced entirely to an eigenvalue
problem for this matrix.

In what follows below we will take all inner products and matrix
multiplications as specified in this subsection.

\section{Perturbed $\mathcal{L}_{_{\mathbf{g}_{-1}}}$-Lorentzian Toda field
theory}

Let us now discuss some theories with the root system enlarged to a
Lorentzian lattice. We start by illustrating the construction principle for
the perturbation of some over extended algebra $\mathbf{g}_{-1}$.

To define these systems we need to enlarge the root space. Adopting the
conventions from \cite{gaberdiel2002class,fring2019n}, the root lattice $%
\Lambda _{\mathbf{g}}$ of the semisimple Lie algebra $\mathbf{g}$ is
extended by a self-dual Lorentzian lattice $\Pi ^{1,1}$ to

\begin{equation}
\Lambda _{\mathbf{g}_{-1}}=\Lambda _{\mathbf{g}}\oplus \Pi ^{1,1}.  \label{L}
\end{equation}%
The root space $\Pi ^{1,1}$ contains two null vectors $k$ and $\bar{k}$ with 
$k\cdot k=\bar{k}\cdot \bar{k}=0$, $k\cdot \bar{k}=1$ and two vectors $\pm
\left( k+\bar{k}\right) $ of length $2$. The simple root system consists in
this case of the $r$ simple roots $\alpha _{1},\ldots ,\alpha _{r}~$of the
semisimple Lie algebra $\mathbf{g}$, the modified affine root $\alpha
_{0}=k-\sum\nolimits_{i=1}^{r}n_{i}\alpha _{i}$, with $n_{i}\in \mathbb{N}$
denoting the Kac labels, and the Lorentzian root $\alpha _{-1}=-\left( k+%
\bar{k}\right) $.

For the corresponding Lagrangians the construction is summarized as follows 
\begin{equation}
\mathcal{L}_{_{\mathbf{g}_{1}}}\overset{\alpha _{0}}{\rightarrow }\mathcal{L}%
_{_{\mathbf{g}_{0}}}\overset{\alpha _{-1}}{\rightarrow }\mathcal{L}_{_{%
\mathbf{g}_{-1}}}\overset{\alpha _{-2}}{\rightarrow }\mathcal{L}_{_{\mathbf{%
\mathring{g}}_{-2}}}.  \label{LLLL}
\end{equation}%
We have started here with the standard conformal Toda field theory $\mathcal{%
L}_{_{\mathbf{g}_{1}}}$ and added the modified root $\alpha _{0}$ to obtain
the massive affine Toda field theory $\mathcal{L}_{_{\mathbf{g}_{0}}}$.
Adding the root $\alpha _{-1}$ yields the scalar field theory described by
the Lagrangian $\mathcal{L}_{_{\mathbf{g}_{-1}}}$, which turns out to be
conformal and can be identified with a theory that is sometimes referred to
as conformal affine Toda field theory \cite%
{babelon1990conformal,aratyn1991kac}. The corresponding algebra is an over
extended algebra, of which for instance $(E_{8})_{-1}$ aka $E_{10}$ is of
relevance in a string/M-theory context \cite{damour200210}. We discuss now
this theory in some detail before we state how the root $\alpha _{-2}$ is
constructed in order to obtain the massive $\mathcal{L}_{_{\mathbf{\mathring{%
g}}_{-2}}}$-theory.

The classical equations of motion for $\mathcal{L}_{_{\mathbf{g}_{-1}}}$,
resulting from $\partial _{\mu }\left[ \partial \mathcal{L}/\partial
(\partial _{\mu }\phi ^{a})\right] =\partial \mathcal{L}/\phi ^{a}$, are 
\begin{equation}
\square \phi ^{a}+\frac{g}{\beta ^{2}}\sum\limits_{i=-1}^{r}\alpha
_{i}^{a}e^{\beta \alpha _{i}\cdot \phi }=0,\qquad a=1,\ldots ,\ell +2.
\label{eqom}
\end{equation}%
We may view the potential in $\mathcal{L}_{_{\mathbf{g}_{-1}}}$as a
perturbation of an affine Toda field theory with potential $\mathcal{V}_{_{%
\mathbf{g}_{0}}}$ so that $\mathcal{V}_{_{\mathbf{g}_{-1}}}=$ $\mathcal{V}%
_{_{\mathbf{g}_{0}}}+\delta \mathcal{V}_{_{\mathbf{g}_{0}}}$, where $\delta 
\mathcal{V}_{_{\mathbf{g}_{0}}}$ corresponds to the term in the sum related
to $\alpha _{-1}$.

Since $\mathcal{V}_{_{\mathbf{g}_{0}}}$ possess a proper vacuum around which
one may expand, it is clear that the additional term $\delta \mathcal{V}_{_{%
\mathbf{g}_{0}}}$ will spoil this property, unless it vanishes by itself for
the value of the vacuum, and the right hand sides in (\ref{eqom}) only
vanish for $\alpha _{i}\cdot \phi \rightarrow -\infty $. An alternative way
to verify whether a theory is conformally invariant or massive is to use the
well-known property of the trace of the improved energy momentum tensor,
that is zero or nonvanishing, respectively. For this purpose we first
transform the equation of motion into a more convenient form by defining a
new field $\Phi _{i}:=\alpha _{i}\cdot \phi -\beta ^{-1}\ln (2\alpha
_{i}^{-2})$, such that the equation of motion (\ref{eqom}) converts into%
\begin{equation}
\square \Phi _{j}+\frac{g}{\beta ^{2}}\sum\limits_{i=-1}^{r}K_{ji}e^{\beta
\Phi _{j}}=0,
\end{equation}%
with $K_{ij}:=2\alpha _{i}\cdot \alpha _{j}/\alpha _{j}^{2}$ denoting the
Cartan matrix, which we assume here to be symmetric, i.e. $\alpha _{j}^{2}=2$%
. Defining further the fields $\varphi _{i}$ through the relations $\Phi
_{i}=\left( M\varphi \right) _{i}$, where the matrix $M$ defined in (\ref{M}%
) factorizes the Cartan matrix as in (\ref{fact}), we obtain the following
version of the equation of motion%
\begin{equation}
\square \varphi _{\alpha }+\frac{g}{\beta ^{2}}\sum\limits_{i=-1}^{r}\left(
M^{T}\right) _{\alpha i}e^{\beta \left( M\varphi \right) _{i}}=0.
\label{equM}
\end{equation}%
Following \cite{curt1982conf,mansfield1983light}, the trace of the improved
energy tensor results to%
\begin{equation}
\Theta _{~\mu }^{\mu }=\sum\limits_{i=-1}^{r}\left( \frac{2g}{\beta ^{2}}%
e^{\beta \left( M\varphi \right) _{i}}+\gamma _{i}\square \varphi
_{i}\right) .
\end{equation}%
Thus for $\gamma _{i}=2\beta ^{-1}\sum_{k}M_{ik}^{-1}$ we obtain the
equation of motion for each term in the sum and the trace of the improved
energy tensor vanishes. In turn, this means when the matrix $M$ is not
invertible the model is not conformally invariant and hence massive.

While $\mathcal{L}_{_{\mathbf{g}_{-1}}}$ is a massless conformally invariant
theory, which might be studied in its own right, here we are interested in
the question of whether it is possible to construct a massive field theory
and therefore add consistently a perturbing term to $\mathcal{V}_{_{\mathbf{g%
}_{-1}}}$%
\begin{equation}
\mathcal{V}_{_{\mathbf{\mathring{g}}_{-2}}}(\phi ):=\mathcal{V}_{_{\mathbf{g}%
_{-1}}}(\phi )+\delta \mathcal{V}_{_{\mathbf{g}_{-1}}}(\phi )=\mathcal{V}_{_{%
\mathbf{g}_{-1}}}(\phi )+\varepsilon \frac{g}{\beta ^{2}}e^{\beta \alpha
_{-2}\cdot \phi }.
\end{equation}%
The vacuum $\phi ^{(0)}$~for the new potential $\mathcal{V}_{_{\mathbf{%
\mathring{g}}_{-2}}}$ computed from the equations $\left. \partial \mathcal{V%
}_{_{\mathbf{\mathring{g}}_{-2}}}/\partial \phi ^{a}\right\vert _{\phi
^{(0)}}=0$, $a=1,\ldots ,r+2$, leads to the constraint%
\begin{equation}
\sum\limits_{i=-1}^{r}\alpha _{i}e^{\beta \alpha _{i}\cdot \phi
^{(0)}}=-\varepsilon \alpha _{-2}e^{\beta \alpha _{-2}\cdot \phi ^{(0)}}.
\end{equation}%
Multiplying with the fundamental weights $\lambda _{j}$ and using the
orthogonality relation $\alpha _{i}\cdot \lambda _{j}=\delta _{ij}$ yields
the relations%
\begin{equation}
e^{\beta \alpha _{i}\cdot \phi ^{(0)}}=-\varepsilon \lambda _{i}\cdot \alpha
_{-2}e^{\beta \alpha _{-2}\cdot \phi ^{(0)}},\qquad i=-1,0,1,\ldots ,r.
\label{id}
\end{equation}%
Expanding now the potential $\mathcal{V}_{_{\mathbf{\mathring{g}}%
_{-2}}}(\phi )$ around the vacuum we obtain with (\ref{id})%
\begin{equation}
\mathcal{V}_{_{\mathbf{\mathring{g}}_{-2}}}(\phi ^{(0)}+\tilde{\phi}%
)=\varepsilon \frac{g}{\beta ^{2}}e^{\beta \alpha _{-2}\cdot \phi ^{(0)}}%
\left[ e^{\beta \alpha _{-2}\cdot \tilde{\phi}}-\sum\limits_{i=-1}^{r}%
\lambda _{i}\cdot \alpha _{-2}e^{\beta \alpha _{i}\cdot \tilde{\phi}}\right]
=\frac{m^{2}}{\beta ^{2}}\sum\limits_{i=-2}^{r}\hat{n}_{i}e^{\beta \alpha
_{i}\cdot \tilde{\phi}},  \label{potential}
\end{equation}%
where $m^{2}=\varepsilon ge^{\beta \alpha _{-2}\cdot \phi ^{(0)}}$, $\hat{n}%
_{-2}=1$ and $\hat{n}_{i}=-\lambda _{i}\cdot \alpha _{-2}$. \ 

We now make the choice $\alpha _{-2}=\bar{k}$, so that with the realizations
of the fundamental weights for $\mathbf{g}_{-1}$ as \cite%
{gaberdiel2002class,fring2019n} 
\begin{equation}
\lambda _{i}=\lambda _{i}^{f}+n_{i}\lambda _{0}^{o},~~\ \lambda _{0}=\bar{k}%
-k,\quad \lambda _{-1}=-k,~~~\text{with }i=1,\ldots ,r,  \label{weights}
\end{equation}%
and $\lambda _{i}$ denoting the fundamental weights of $\mathbf{g}$, we
compute $\hat{n}_{-1}=1$, $\hat{n}_{0}=1$ and $\hat{n}_{i}=n_{i}$. Notice
that $\mathbf{\mathring{g}}_{-2}$ is not a proper over extended algebra as
defined in \cite{fring2019n}, hence the notation $\mathbf{\mathring{g}}_{-2}$
instead of $\mathbf{g}_{-2}$. We have $\alpha _{-2}\in $ $\Lambda _{\mathbf{g%
}}\oplus \Pi ^{1,1}$ connecting in an almost identical way as the root $%
k-\left( \ell +\bar{\ell}\right) $ to all the other simple roots with $%
\alpha _{-2}\cdot \alpha _{-1}=-1$, $\alpha _{-2}\cdot \alpha _{i}=0$, $%
i=1,\ldots ,r$. However, this root also connects to the affine root $\alpha
_{-2}\cdot \alpha _{0}=1$, has length zero, i.e. $\alpha _{-2}^{2}=0\neq 2$,
and is defined in a smaller representation space than the standard $\alpha
_{-2}$-root. Hence $\Lambda _{\mathbf{\mathring{g}}_{-2}}$ can not be viewed
as a lattice related to a Kac-Moody algebra and we refer to it therefore as
a root lattice to an \emph{almost over extended algebra}.

Expanding now (\ref{potential}) around zero we obtain a constant term in the
potential of the form $m^{2}/\beta ^{2}(\hat{n}_{-2}+\hat{n}_{-1}+\hat{n}%
_{0}+\sum\nolimits_{i=1}^{r}n_{i}$ $)=m^{2}/\beta ^{2}(2+h)$ with $h$
denoting the Coxeter number of $\mathbf{g}$. Crucially, our choice for $%
\alpha _{-2}$ also has the desired property that the linear term in the
expansion vanishes, because $\sum\nolimits_{i=-2}^{r}\hat{n}_{i}\alpha
_{i}=0 $. Labeling rows and columns as $(\tilde{\phi}_{1},\ldots ,\tilde{\phi%
}_{r+2})$ the square mass matrix is obtained as%
\begin{equation}
M^{2}=m^{2}\sum\limits_{i=-2}^{r}\hat{n}_{i}\left( 
\begin{array}{ccccc}
\alpha _{i}^{1}\alpha _{i}^{1} & \ldots & \alpha _{i}^{1}\alpha _{i}^{r} & 
-\alpha _{i}^{1}\alpha _{i}^{r+2} & -\alpha _{i}^{1}\alpha _{i}^{r+1} \\ 
\vdots & \ddots & \vdots & \vdots & \vdots \\ 
\alpha _{i}^{1}\alpha _{i}^{r} & \ldots & \alpha _{i}^{r}\alpha _{i}^{r} & 
-\alpha _{i}^{r}\alpha _{i}^{r+2} & -\alpha _{i}^{r}\alpha _{i}^{r+1} \\ 
-\alpha _{i}^{1}\alpha _{i}^{r+2} & \ldots & -\alpha _{i}^{r}\alpha
_{i}^{r+2} & \alpha _{i}^{r+2}\alpha _{i}^{r+2} & \alpha _{i}^{r+1}\alpha
_{i}^{r+2} \\ 
-\alpha _{i}^{1}\alpha _{i}^{r+1} & \ldots & -\alpha _{i}^{r}\alpha
_{i}^{r+1} & \alpha _{i}^{r+1}\alpha _{i}^{r+2} & \alpha _{i}^{r+1}\alpha
_{i}^{r+1}%
\end{array}%
\right) .  \label{M2}
\end{equation}%
The classical mass spectra resulting from (\ref{M2}) are only physically
meaningful when the eigenvalues of $M^{2}$ are real and positive. Before we
will discuss concrete examples below, we first establish whether these type
of theories are integrable by performing the Painlev\'{e} integrability test.

\section{Painlev\'{e} integrability test}

We now largely follow the line of reasoning in \cite%
{yoshida1983,gebert1996toda} and generalize the Painlev\'{e} test \cite%
{Painor,ARS,Pain1,Joshi1,Gramma} in order to establish whether variations of
the $\mathcal{L}_{_{\mathbf{g}_{-n}}}$ Lorentzian Toda theories and
perturbations thereof are integrable. First we transform the equation of
motion in version (\ref{equM}) into light-cone coordinates so that $\square
=\partial _{-}\partial _{+}$. For the sake of brevity, we denote $\partial
_{-}$ by an overdot and $\partial _{+}$ by an overdash, e.g. $\partial
_{-}\varphi =:\dot{\varphi}$ and $\partial _{+}\varphi =:\acute{\varphi}$.
For further convenience we set $g=\beta =1$. We start by separating the
second order equation of motion into two two first order equations, which
can be achieved by introduce two quantities, akin but not equal to canonical
variables, as 
\begin{equation}
P_{\alpha }=\Dot{\varphi _{\alpha }},\qquad Q_{i}=e^{\left( M\varphi \right)
_{i}},~~\ \ \ \ \ \alpha =1,\ldots ,\ell +2m\text{,}i=1,\ldots ,r+n+1.
\end{equation}%
Differentiating these quantities with respect to each light-cone coordinate
we obtain%
\begin{equation}
\acute{P}_{\alpha }=\square \varphi _{\alpha }=-\sum\limits_{i=-n}^{r}\left(
M^{T}\right) _{\alpha i}Q_{i},~~~~~~~\dot{Q}_{i}=Q_{i}\left( MP\right) _{i}.
\label{eqm}
\end{equation}%
We now Painlev\'{e} expand $P_{\alpha }$ and $Q_{i}$, making the standard
assumption that both quantities possess movable critical singularities in
some field $\phi (x_{-},x_{+})\rightarrow 0$, whose leading order is
determined by some positive integers $n_{p},n_{q}>0$

\begin{equation}
Q_{i}=\sum_{k=0}^{\infty }a_{i}^{(k)}\phi ^{k-n_{q}},\hspace{4mm}%
~~~P_{\alpha }=\sum_{k=0}^{\infty }b_{\alpha }^{(k)}\phi ^{k-n_{p}}.
\label{Laurent}
\end{equation}

\noindent Differentiating the expansions we obtain

\begin{equation}
\dot{Q}_{i}=\sum_{k=0}^{\infty }(k-n_{q})a_{i}^{(k)}\phi ^{k-n_{q}-1}\Dot{%
\phi}\hspace{1mm},~~~~~\ \acute{P}_{\alpha }=\sum_{k=0}^{\infty
}(k-n_{p})b_{\alpha }^{(k)}\phi ^{k-n_{p}-1}\acute{\phi}.  \label{PQexpanded}
\end{equation}%
Substituting next the expansions (\ref{Laurent}) and (\ref{PQexpanded}) into
(\ref{eqm}) and balancing the powers we obtain%
\begin{eqnarray}
(k-n_{p})\acute{\phi}b_{\alpha }^{(k)} &=&-\sum\limits_{i=-n}^{r}\left(
M^{T}\right) _{\alpha i}a_{i}^{(k)},  \label{e1} \\
(k-n_{q})\Dot{\phi}a_{i}^{(k)} &=&\sum_{m=0}^{k}a_{i}^{(k-m)}\left(
Mb^{(m)}\right) _{i},  \label{e2}
\end{eqnarray}%
with $n_{q}=n_{p}+1$. At this point we have to distinguish between two cases
i) when the Cartan matrix is invertible and ii) when it is not.

\subsection{Invertible Cartan matrix}

For $k=0$ we can solve the equations (\ref{e1}) and (\ref{e2}) for the
leading order coefficient functions when the Cartan matrix is invertible%
\begin{equation}
a_{i}^{(0)}=-n_{p}n_{q}\Dot{\phi}\acute{\phi}D_{i},~~\ ~~b_{\alpha
}^{(0)}=-n_{q}\Dot{\phi}\sum\limits_{i=-n}^{r}\left( M^{T}\right) _{\alpha
i}D_{i},  \label{e3}
\end{equation}%
where the $D_{i}$ are the constants as defined in (\ref{Di}).

Next we extract in (\ref{e2}) the terms in the sum for $m=0$ and $m=k$.
Using also $\left( Mb^{(0)}\right) _{i}=-n_{q}\Dot{\phi}$, we re-write (\ref%
{e1}) and (\ref{e2}) as%
\begin{eqnarray}
k\Dot{\phi}a_{i}^{(k)}+n_{p}n_{q}\Dot{\phi}\acute{\phi}D_{i}\left(
Mb^{(k)}\right) _{i} &=&\sum_{m=1}^{k-1}a_{i}^{(k-m)}\left( Mb^{(m)}\right)
_{i},  \label{s1} \\
\sum\limits_{i=-n}^{r}\left( M^{T}\right) _{\alpha i}a_{i}^{(k)}+(k-n_{p})%
\acute{\phi}b_{\alpha }^{(k)} &=&0.  \label{s2}
\end{eqnarray}

\noindent These equations, (\ref{s1}) and (\ref{s2}), can be converted into
matrix form \noindent

\begin{equation}
T^{(k)}X^{(k)}=Y^{(k)},  \label{central}
\end{equation}

\noindent when defining the $N+M=(r+n+1)+(\ell +2m)$ dimensional column
vectors

\begin{eqnarray}
X^{(k)} &=&(a_{1}^{(k)},\cdots ,a_{M}^{(k)},b_{1}^{(k)},\cdots
,b_{N}^{(k)})^{T}, \\
Y^{(k)} &=&\sum_{m=1}^{k-1}(a_{1}^{(k-m)}\left( Mb^{(m)}\right) _{1},\cdots
,a_{M}^{(k-m)}\left( Mb^{(m)}\right) _{M},0,\cdots ,0)^{T},
\end{eqnarray}

\noindent together with the $(M+N)\times (M+N)$-matrix

\begin{equation}
T^{(k)}=\left( 
\begin{array}{ll}
A_{M\times M}^{(k)} & B_{M\times N}^{(k)} \\ 
C_{N\times M}^{(k)} & E_{N\times N}^{(k)}%
\end{array}%
\right) .
\end{equation}%
The block matrices in $T$ have entries%
\begin{equation}
A_{ij}^{(k)}=k\Dot{\phi}\delta _{ij},\quad B_{i\alpha }^{(k)}=n_{p}n_{q}\Dot{%
\phi}\acute{\phi}D_{i}M_{i\alpha },\quad C_{\alpha i}^{(k)}=\left(
M^{T}\right) _{\alpha i},\quad E_{\alpha \beta }^{(k)}=(k-n_{p})\acute{\phi}%
\delta _{\alpha \beta }.
\end{equation}%
Equation (\ref{central}) is the central equation for the Painlev\'{e}
integrability test. It is a recursive equation that may in principle be
solved iteratively at each level $k$ for the coefficient functions contained
in $X^{(k)}$ as long as the matrix $T^{(k)}$ is invertible. Whenever this is
not the case one is introducing a free parameter, a \emph{resonance} in
Painlev\'{e} integrability test parlance, into the set of equations. When
there are enough resonances in the system as boundary conditions or
integration constants, the system is passing the test and is said to be
integrable.

Let us therefore compute the determinant of $T^{(k)}$. Using the identity%
\begin{equation}
\det \left( 
\begin{array}{ll}
A & B \\ 
C & E%
\end{array}%
\right) =\det \left( 
\begin{array}{ll}
A & B \\ 
C & E%
\end{array}%
\right) \det \left( 
\begin{array}{cc}
I & 0 \\ 
-E^{-1}C & I%
\end{array}%
\right) =\det (A-BE^{-1}C)\det (E),
\end{equation}%
we obtain%
\begin{equation}
\det T^{(k)}=(k-n_{p})^{r+n}\acute{\phi}^{r+n+1}\Dot{\phi}\det \left[
k(k-n_{p})I-n_{p}(n_{p}+1)DK\right] .  \label{dt}
\end{equation}%
Apart from the pre-factor, for $n=n_{p}=1$ this reduces to the expression
previously obtained in \cite{gebert1996toda} for the hyperbolic Kac-Moody
algebras. Taking now $n_{p}=1$, the matrix in the determinant becomes the 
\emph{Painlev\'{e} matrix} and the last factor in (\ref{dt}) can be read as
the characteristic equation for the matrix $P=2DK$ with eigenvalues $k(k-1)$%
. Thus we have found that also for the $\mathcal{L}_{_{\mathbf{g}_{-n}}}$
Lorentzian Toda theories the integrability test can be reduced to an
eigenvalue problem for $P$. Nicolai and Olive noticed in \cite%
{nicolai2001dio} that this matrix also emerges from the adjoint action of
the $SO(1,2)$ Casimir operator on the Cartan subalgebra and that in fact the
eigenvalues are identical to the Casimir eigenvalues. In this generalised
case a principal $SO(1,2)$-subalgebra does not always exist, as explicitly
argued in \cite{fring2019n} for many cases, so that it needs to be replaced
in part by a principal $SO(3)$-subalgebra.

\subsection{Non-invertible Cartan matrix}

When the Cartan matrix is not invertible we can not derive (\ref{e3}) from
the equations (\ref{e1}) and (\ref{e2}). As a specific theory that involves
a non-invertible Cartan matrix let us know consider the $\mathcal{L}_{_{%
\mathbf{g}_{0}}}$-theory, corresponding to affine Toda theory. Of course in
this case we know that the theory is integrable since exact Lax pairs have
been constructed for the classical theory \cite{Integ} and in the quantum
case the S-matrix factorizes into two-particle S-matrices as a consequence
of the integrability \cite{BCDS}. However, let us see how the Painlev\'{e}
test can be implemented, since the same line of argumentation can then also
be applied to some extended theories we consider below. Using the fact that $%
K_{ij}=\tilde{K}_{ij}$ for $i,j=1,\ldots ,r$ with $\tilde{K}$ denoting the
invertible Cartan matrix of $\mathbf{g}$, we can split off the last row and
the last column from $K$. Then it is easily seen that (\ref{e3}) is replaced
by%
\begin{equation}
a_{i}^{(0)}=-n_{p}n_{q}\Dot{\phi}\acute{\phi}\tilde{D}%
_{i}+n_{i}a_{0}^{(0)},~~\ ~~b_{\alpha }^{(0)}=-n_{q}\Dot{\phi}%
\sum\limits_{i=1}^{r}\left( M^{T}\right) _{\alpha i}\tilde{D}_{i},
\end{equation}%
where $\tilde{D}_{i}:=\sum\nolimits_{j=1}^{r}\tilde{K}_{ij}^{-1}$ and the $%
n_{i}$ denote the Kac labels as defined after (\ref{L}). Following now the
same steps as in the previous subsection we derive the matrix $T$ with block
matrices%
\begin{equation}
A_{ij}^{(k)}=k\Dot{\phi}\delta _{ij},\quad B_{i\alpha }^{(k)}=n_{p}n_{q}\Dot{%
\phi}\acute{\phi}D_{i}M_{i\alpha }-n_{i}M_{i\alpha }a_{0}^{(0)},\quad
C_{\alpha i}^{(k)}=\left( M^{T}\right) _{\alpha i},\quad E_{\alpha \beta
}^{(k)}=(k-n_{p})\acute{\phi}\delta _{\alpha \beta },
\end{equation}%
where we defined $D_{0}:=0$. Taking now $a_{0}^{(0)}=0$, we notice that the
only non-vanishing entry in the $0$-row of $T^{(k)}$ is $%
T_{00}^{(k)}=A_{00}^{(k)}=k\Dot{\phi}$. We can then expand $\det T^{(k)}$
with respect to the first row and derive%
\begin{equation}
\det T^{(k)}=k(k-n_{p})^{r+1}\acute{\phi}^{r+2}\Dot{\phi}^{2}\det \left[
k(k-n_{p})I_{r\times r}-n_{p}(n_{p}+1)\tilde{D}\tilde{K}\right] ,
\end{equation}%
with $\tilde{D}$, $\tilde{K}$ belonging to $\mathbf{g}$. Thus we have
reduced the Painlev\'{e} test for the $\mathcal{L}_{_{\mathbf{g}_{0}}}$%
-theory to an eigenvalue problem for the matrix $n_{p}(n_{p}+1)\tilde{D}%
\tilde{K}$ associated to $\mathbf{g}$.

Thus we conclude that the integrability properties of the $\mathcal{L}_{_{%
\mathbf{g}}}$-theory are inherited by the $\mathcal{L}_{_{\mathbf{g}_{0}}}$%
-theory, that is when $\mathcal{L}_{_{\mathbf{g}}}$ is (non)integrable so is 
$\mathcal{L}_{_{\mathbf{g}_{0}}}$.

For simplicity we derived here the eigenvalue equation (\ref{dt}) for
symmetric Cartan matrices. We may repeat the same line of argumentation by
replacing in $M^{T}$ roots by coroots, $\alpha _{i}\rightarrow \hat{\alpha}%
_{i}=2\alpha _{i}/\alpha _{i}^{2}$ when $\alpha _{i}^{2}\neq 0$. Then it is
easily seen that (\ref{dt}) generalizes to the nonsymmetric case for which
the Cartan matrix is defined as $K_{ij}=2\alpha _{i}\cdot \alpha _{j}/\alpha
_{j}^{2}$ when $\alpha _{j}^{2}\neq 0$ and remains $K_{ij}=\alpha _{i}\cdot
\alpha _{j}$ when $\alpha _{j}^{2}=0$.

\section{The characteristic equation of the Painlev\'{e} matrix}

We will keep now $n_{p}=1$ and analyse the characteristic equation for the
Painlev\'{e} matrix $P$ as defined in (\ref{P}) 
\begin{equation}
\det \left[ k(k-1)I-P\right] =0,
\end{equation}%
in some more detail. As argued in the previous subsection, for any version
of the Lorentzian Toda field theories to be integrable the eigenvalues of
the Painlev\'{e} matrix must be integer valued and factorize as $k(k-1)$
with $k\in \mathbb{N}$. In particular, this means when the eigenvalues are
negative the theory is not integrable. These cases can be identified easily.
We need to argue differently depending on whether the matrix $D$ is positive
or negative definite, semi-definite of indefinite:

Denoting by $\limfunc{ind}A=e_{p}-e_{n}$ the \emph{index of the matrix} $A$,
defined as the difference between the positive and negative eigenvalues of $%
A $, $e_{p}$ and $e_{n}$, respectively, we have the relation%
\begin{equation}
\limfunc{ind}(\pm 2DK)=\limfunc{ind}(K),  \label{indk}
\end{equation}%
where the $+$sign holds for $D$ positive definite and the $-$sign for $D$
negative definite.

To prove this relation we first note that the matrix $\sqrt{\pm D}K\sqrt{\pm
D}$ has the same eigenvalues as $\pm DK$. Here $\sqrt{\pm D}$ is the
positive square root with the sign depending on whether $D$ is positive or
negative definite. Next we invoke Sylvester's theorem, see e.g. Theorem 12.3
in \cite{carrell2005}, which states that two symmetric square matrices $A$
and $B$ that are congruent to each other, i.e. $A=QBQ^{T}$ for some
nonsingular matrix $Q$, have the same index. Applied to the above this means
that $\limfunc{ind}(\sqrt{\pm D}K\sqrt{\pm D})=\limfunc{ind}(K)$, since $%
\sqrt{\pm D}^{T}=\sqrt{\pm D}$. Therefore with $\limfunc{ind}(\sqrt{\pm D}K%
\sqrt{\pm D})=\limfunc{ind}(\pm DK)$ we obtain (\ref{indk}).

When $D$ is semi-definite we can define a reduced $D$-matrix as $\hat{D}$ by
setting the positive or negative entries to zero and use a reduced version
of (\ref{indk}) as $\limfunc{ind}(\pm 2\hat{D}K)=\limfunc{ind}(K)$.

Since a necessary condition for passing the Painlev\'{e} test is that all
eigenvalues of $2DK$ are positive, i.e. $\limfunc{ind}(2DK)=\ell $ with $%
\ell $ denoting the rank of $K$, the relation (\ref{indk}) implies that $%
\limfunc{ind}(\pm K)=\ell \,$. This means only Lorentzian Toda field
theories based on positive or negative definite Cartan matrix can pass the
Painlev\'{e} test. In turn this means that those theories build from
non-definite Cartan matrices can not be integrable.

\section{Constructions of Lorentzian Toda field theory}

We will now construct various types of Toda field theories based on
different versions of root systems corresponding to Lorentzian Kac-Moody
algebras and their extensions. We will encounter conformally invariant and
massive models.

\subsection{$\mathcal{L}_{_{\mathbf{\mathring{g}}_{-n}}}$-extended
Lorentzian Toda field theory}

This first type of theories is a series constituting an infinite extension
of the perturbed $\mathcal{L}_{_{\mathbf{g}_{-1}}}$-theory introduced in
section 3.1. The theories in this series come in two variants: The $\mathcal{%
L}_{_{\mathbf{\mathring{g}}_{-n}}}$-Lorentzian Toda field theories for odd $%
n $ are conformally invariant and those for which $n$ is even are massive.
As a construction principle we extend the one previously used for the
perturbation of the $\mathcal{L}_{_{\mathbf{g}_{-1}}}$-theory and build the
roots as follows. For the massless $\mathcal{L}_{_{\mathbf{\mathring{g}}%
_{-(2n-1)}}}$-theories we have the $r+2n$ roots%
\begin{equation}
\begin{array}{ll}
\alpha _{i}\equiv \text{simple roots of }\mathbf{g} & \text{for }j=1,\ldots
,r, \\ 
\alpha _{-(2i-2)}=k_{i}-\sum\nolimits_{j=-(2i-3)}^{r}n_{j}\alpha _{j}~~~\ \
\  & \text{for }i=1,\ldots ,n, \\ 
\alpha _{-(2i-1)}=-(k_{i}+\bar{k}_{i}) & \text{for }i=1,\ldots ,n.%
\end{array}
\label{roots1}
\end{equation}%
We notice that the roots $\alpha _{-(2i-2)}$ have length zero for $%
i=2,\ldots ,n$, have a standard inner product equal to $-1$ with nearest
neighbour roots on the Dynkin diagram and a more unusual inner product equal
to $1$ for next to nearest neighbours. The roots $\alpha _{-(2i-1)}$ have
length $2$ for $i=1,\ldots ,n$. Thus we have the inner products 
\begin{equation}
\alpha _{-(2i-2)}^{2}=0,\quad \alpha _{-(2j-1)}^{2}=2,\quad \alpha
_{-k}\cdot \alpha _{-(k+1)}=-1,\quad \alpha _{-2l}\cdot \alpha _{-(2l+2)}=1,
\end{equation}%
for $i=2,\ldots ,n$, $j=1,\ldots ,n$, $k=1,\ldots ,2n-2$ and $l=0,1,\ldots
,n-2$. At each affine root $\alpha _{0}$ the Dynkin diagram is extended by
the following segment:

\setlength{\unitlength}{0.58cm} 
\begin{picture}(14.00,3.5)(-1.0,7.5)
\thicklines
\put(0.5,9.0){\Large{$\mathbf{\mathring{g}}_{1-2n}:$}}
\put(3.4,9.2){\Large{$\ldots$}}
\put(4.5,9.2){\line(1,0){0.6}}
\qbezier[18](5.3,9.4)(6.2,10.5)(7.2,9.4)
\qbezier[18](7.3,9.4)(8.2,10.5)(9.2,9.4)
\qbezier[18](12.3,9.4)(13.2,10.5)(14.2,9.4)
\put(5.0,9.0){\Large{$\bullet$}}
\put(5.4,9.2){\line(1,0){0.9}}
\put(4.9,8.5){{$\alpha_0$}}
\put(6.0,9){\Large{$\bullet$}}
\put(6.2,9.2){\line(1,0){0.9}}
\put(5.8,8.5){{$\alpha_{-1}$}}
\put(7.0,9){\Large{$\circ$}}
\put(7.4,9.2){\line(1,0){0.9}}
\put(6.9,8.5){{$\alpha_{-2}$}}
\put(8.0,9){\Large{$\bullet$}}
\put(8.2,9.2){\line(1,0){0.9}}
\put(8.0,8.5){{$\alpha_{-3}$}}
\put(9.0,9){\Large{$\circ$}}
\put(9.4,9.2){\line(1,0){0.7}}
\put(9.2,8.5){{$\alpha_{-4}$}}
\put(10.2,9.2){\Large{$\ldots$}}
\put(11.4,9.2){\line(1,0){0.7}}
\put(12.0,9){\Large{$\circ$}}
\put(12.4,9.2){\line(1,0){0.9}}
\put(11.6,8.5){{$\small{\alpha_{4-2n}}$}}
\put(13.0,9){\Large{$\bullet$}}
\put(13.2,9.2){\line(1,0){0.9}}
\put(14.0,9){\Large{$\circ$}}
\put(14.4,9.2){\line(1,0){0.9}}
\put(13.6,8.5){{$\small{\alpha_{2-2n}}$}}
\put(15.0,9){\Large{$\bullet$}}
\end{picture}

We used here the standard conventions for drawing Dynkin diagrams related to
semi-simple Lie algebras in which vertices with bullets indicate roots of
length $2$ and single line links between two vertices correspond to inner
products of $-1$ between the two corresponding roots. We increase the set of
rules by indicating roots of length $0$ with an empty circles and inner
products of $1$ by dotted links between two vertices correspond to the
roots. Such type of zero length roots and inner products equal to $1$ are
not entirely unusual and also occur in the context of Lie superalgebras and
of their affine extensions \cite{frappat1989structure}.

The corresponding Cartan matrix is 
\begin{equation}
K_{\mathbf{\mathring{g}}_{-(2n-1)}}=\left( 
\begin{array}{llllccc}
&  &  & q_{1} & 0 & \cdots & 0 \\ 
& K_{\mathbf{g}} &  & \vdots & \vdots &  & \vdots \\ 
&  &  & q_{r} & 0 & \cdots & 0 \\ 
q_{1} & \cdots & q_{r} & 2 & \multicolumn{1}{l}{-1} & \multicolumn{1}{l}{1}
& \multicolumn{1}{l}{0} \\ 
\multicolumn{1}{c}{0} & \multicolumn{1}{c}{\cdots} & \multicolumn{1}{c}{0} & 
-1 & \multicolumn{1}{l}{} & \multicolumn{1}{l}{} & \multicolumn{1}{l}{} \\ 
\multicolumn{1}{c}{\vdots} & \multicolumn{1}{c}{} & \multicolumn{1}{c}{\vdots
} & 1 & \multicolumn{1}{l}{} & \multicolumn{1}{l}{\hat{K}_{2n-1}} & 
\multicolumn{1}{l}{} \\ 
\multicolumn{1}{c}{0} & \multicolumn{1}{c}{\cdots} & \multicolumn{1}{c}{0} & 
0 & \multicolumn{1}{l}{} & \multicolumn{1}{l}{} & \multicolumn{1}{l}{}%
\end{array}%
\right)  \label{Kn1}
\end{equation}%
with $q_{s}:=\alpha _{0}\cdot \alpha _{s}$, $s=1,\ldots ,r$, and $%
(2n-1)\times (2n-1)$ matrix $\hat{K}_{2n-1}$ with entries%
\begin{equation}
\hat{K}_{2i-1,2i-1}=2,\quad \hat{K}_{2j,2j}=0,\quad \hat{K}_{k,k+1}=-1,\quad 
\hat{K}_{2l,2l+2}=1,  \label{KK}
\end{equation}%
for $i=1,\ldots ,n$, $j=1,\ldots ,n-1$, $k=1,\ldots ,2n-2$ and $l=1,\ldots
,n-2$.

Taking the same roots and adding one root at the end of the Dynkin diagram
as the negative highest root, designed to make the linear term in the
potential vanish, we obtain the massive $\mathcal{\mathring{L}}_{_{\mathbf{g}%
_{-(2n)}}}$-theory based on $r+2n+1$ roots%
\begin{equation}
\begin{array}{ll}
\alpha _{i}\equiv \text{simple roots of }\mathbf{g} & \text{for }j=1,\ldots
,r, \\ 
\alpha _{-(2i-2)}=k_{i}-\sum\nolimits_{j=-(2i-3)}^{r}n_{j}\alpha _{j}~~~\ \
\  & \text{for }i=1,\ldots ,n, \\ 
\alpha _{-(2i-1)}=-(k_{i}+\bar{k}_{i}) & \text{for }i=1,\ldots ,n, \\ 
\alpha _{-(2n)}=-\sum\nolimits_{j=-(2n-1)}^{r}n_{j}\alpha _{j}. & 
\end{array}
\label{roots2}
\end{equation}%
Now at each affine root $\alpha _{0}$ the Dynkin diagram is extended by the
segment:

\setlength{\unitlength}{0.58cm} 
\begin{picture}(14.00,3.5)(-1.0,3.5)
\thicklines
\put(0.5,5.0){\Large{$\mathbf{\mathring{g}}_{-2n}:$}}
\put(3.4,5.2){\Large{$\ldots$}}
\put(4.5,5.2){\line(1,0){0.6}}
\qbezier[18](5.3,5.4)(6.2,6.5)(7.2,5.4)
\qbezier[18](7.3,5.4)(8.2,6.5)(9.2,5.4)
\qbezier[18](12.3,5.4)(13.2,6.5)(14.2,5.4)
\qbezier[18](14.3,5.4)(15.2,6.5)(16.2,5.4)
\put(5.0,5){\Large{$\bullet$}}
\put(5.4,5.2){\line(1,0){0.9}}
\put(4.8,4.5){{$\alpha_0$}}
\put(6.0,5){\Large{$\bullet$}}
\put(6.2,5.2){\line(1,0){0.9}}
\put(5.8,4.5){{$\alpha_{-1}$}}
\put(7.0,5){\Large{$\circ$}}
\put(7.4,5.2){\line(1,0){0.9}}
\put(6.9,4.5){{$\alpha_{-2}$}}
\put(8.0,5){\Large{$\bullet$}}
\put(8.2,5.2){\line(1,0){0.9}}
\put(8.0,4.5){{$\alpha_{-3}$}}
\put(9.0,5){\Large{$\circ$}}
\put(9.4,5.2){\line(1,0){0.7}}
\put(9.2,4.5){{$\alpha_{-4}$}}
\put(10.2,5.2){\Large{$\ldots$}}
\put(11.4,5.2){\line(1,0){0.7}}
\put(12.0,5){\Large{$\circ$}}
\put(12.4,5.2){\line(1,0){0.9}}
\put(11.6,4.5){{$\small{\alpha_{4-2n}}$}}
\put(13.0,5){\Large{$\bullet$}}
\put(13.2,5.2){\line(1,0){0.9}}
\put(14.0,5){\Large{$\circ$}}
\put(14.4,5.2){\line(1,0){0.9}}
\put(13.6,4.5){{$\small{\alpha_{2-2n}}$}}
\put(15.0,5){\Large{$\bullet$}}
\put(15.2,5.2){\line(1,0){0.9}}
\put(16.0,5){\Large{$\circ$}}
\put(15.6,4.5){{$\small{\alpha_{-2n}}$}}
\end{picture}

\noindent The corresponding Cartan matrix is 
\begin{equation}
K_{\mathbf{\mathring{g}}_{-2n}}=\left( 
\begin{array}{ccccccc}
&  &  & q_{1} & 0 & \cdots & 0 \\ 
& K_{\mathbf{g}} &  & \vdots & \vdots &  & \vdots \\ 
&  &  & q_{r} & 0 & \cdots & 0 \\ 
q_{1} & \cdots & q_{r} & 2 & -1 & 1 & 0 \\ 
0 & \cdots & 0 & -1 &  &  &  \\ 
\vdots &  & \vdots & 1 &  & \hat{K}_{2n} &  \\ 
0 & \cdots & 0 & 0 &  &  & 
\end{array}%
\right)
\end{equation}%
where the entries of the $(2n)\times (2n)$ matrix $\hat{K}_{2n}$ are defined
as in (\ref{KK}) with $i=1,\ldots ,n$, $j=1,\ldots ,n$, $k=1,\ldots ,2n-1$
and $l=1,\ldots ,n-1$.

For the Toda field theories constructed from these root systems it follows
from section 4 and 5 that the Painlev\'{e} integrability test is entirely
reduced to an eigenvalue problem for the Painlev\'{e} matrix $P$, which must
factor as $n(n-1)$ with $n$ being an integer. For the semi-simple Lie
algebras these integer have been identified as the exponents related to
properties of the Casimir operator of the principle subalgebra on one hand 
\cite{nicolai2001dio} and on the other as labeling the spins of conserved
W-algebra currents \cite{balog1990toda}. From the arguments in section 4.2
it also follows directly that we can reduce the test to the $\mathcal{L}_{_{%
\mathbf{\mathring{g}}_{-2n}}}$-extended Lorentzian Toda field theory to the
eigenvalue problem for $2D_{\mathbf{\mathring{g}}_{-(2n-1)}}K_{\mathbf{%
\mathring{g}}_{-(2n-1)}}$.

Let us now study these theories for some concrete algebras in more detail.

\subsubsection{$\mathbf{(\mathring{A}}_{2}\mathbf{)}_{-2}$-Lorentzian Toda
field theories}

We start with a simply system, the $\mathbf{(\mathring{A}}_{2}\mathbf{)}%
_{-2} $-Lorentzian Toda field theories. We represent the $\mathbf{(\mathring{%
A}}_{2}\mathbf{)}_{-2}$ roots (\ref{roots2}) on a four dimensional lattice as%
\begin{eqnarray}
\alpha _{1} &=&\left( \sqrt{\frac{3}{2}},-\sqrt{\frac{1}{2}};0,0\right)
,\quad \alpha _{2}=\left( 0,\sqrt{2};0,0\right) ,\quad \alpha _{0}=\left( -%
\sqrt{\frac{3}{2}},\sqrt{\frac{1}{2}}-\sqrt{2};1,0\right) ,~~\ \ ~~ \\
\alpha _{-1} &=&\left( 0,0;-1,1\right) ,\quad \alpha _{-2}=\left(
0,0;0,-1\right) .
\end{eqnarray}%
The analogue of the affine root is $\alpha
_{-2}=-\sum\nolimits_{j=-1}^{2}n_{j}\alpha _{j}$ with all Kac labels $%
n_{j}=1 $. It is easily checked that indeed the roots $\alpha _{-1},\alpha
_{0},\alpha _{1},\alpha _{2}$ have length $2$ and the root $\alpha _{-2}$
has length $0$. The Dynkin diagram drawn with the standard rules augmented
with the set of rules as stated at the end of the previous subsection is
therefore:

\setlength{\unitlength}{0.58cm} 
\begin{picture}(14.00,3.5)(-1.0,4)
\thicklines
\put(1.5,5.0){\Large{$\mathbf{\mathring{A}}_{-2}:$}}
\qbezier[18](7.3,5.4)(8.2,6.5)(9.2,5.4)
\put(5.3,6){{$\small{\alpha_{1}}$}}
\put(5.3,4){{$\small{\alpha_{2}}$}}
\put(6.0,6){\Large{$\bullet$}}
\put(6.0,4){\Large{$\bullet$}}
\put(7.2,5.2){\line(-1,1){0.9}}
\put(7.2,5.2){\line(-1,-1){0.9}}
\put(7.0,5){\Large{$\bullet$}}
\put(7.4,5.2){\line(1,0){0.9}}
\put(6.8,4.5){{$\small{\alpha_{0}}$}}
\put(8.0,5){\Large{$\bullet$}}
\put(8.2,5.2){\line(1,0){0.9}}
\put(7.8,4.5){{$\small{\alpha_{-1}}$}}
\put(9.0,5){\Large{$\circ$}}
\put(9.0,4.5){{$\small{\alpha_{-2}}$}}
\end{picture}

The eigenvalues of the Cartan matrix $K_{\mathbf{(\mathring{A}}_{2}\mathbf{)}%
_{-1}}$ are $(3.48119,3.,1.68889,-0.170086)$, with exactly one negative
eigenvalue as we expect. The mass matrix (\ref{M2}) for this root system is
computed to 
\begin{equation}
M^{2}=\frac{1}{2}m^{2}\left( 
\begin{array}{cccc}
3 & 0 & 0 & \sqrt{\frac{3}{2}} \\ 
0 & 3 & 0 & \frac{1}{\sqrt{2}} \\ 
0 & 0 & 2 & -1 \\ 
\sqrt{\frac{3}{2}} & \frac{1}{\sqrt{2}} & -1 & 2%
\end{array}%
\right) ,
\end{equation}%
with positive, that is physical, eigenvalues $(4.1701,3,2.3111,0.51880)$ for 
$m=\sqrt{2}$. The matrix $D_{\mathbf{(\mathring{A}}_{2}\mathbf{)}_{-1}}$ as
defined in (\ref{Di}) is negative definite with $D_{1}=D_{2}=-6$, $D_{3}=-7$
and $D_{4}=-3$. The eigenvalues of the Painlev\'{e} matrix $P$ are $%
(-42,-36,-12,2)$ and the relation (\ref{indk}) is confirmed as $\limfunc{ind}%
(-2DK)=\limfunc{ind}(K)=2$. The theory fails the Painlev\'{e} test and is
therefore not integrable.

\subsubsection{$(\mathbf{\mathring{E}}_{8}\mathbf{)}_{-2n}$-Lorentzian Toda
field theories}

The first member of the $(\mathbf{\mathring{E}}_{8}\mathbf{)}_{-2n}$-series
is the $(\mathbf{\mathring{E}}_{8}\mathbf{)}_{0}$-theory corresponding to
the well studied affine Toda field theories, that describes the scaling
limit of the Ising model at critical temperature in magnetic field \cite%
{zamo1989}. The next member is the $(\mathbf{\mathring{E}}_{8}\mathbf{)}%
_{-2} $-theory for which we represent the roots (\ref{roots2}) on a ten
dimensional root lattice as%
\begin{equation}
\begin{array}{ll}
\alpha _{1}=\left( \frac{1}{2},-\frac{1}{2},-\frac{1}{2},-\frac{1}{2},-\frac{%
1}{2},-\frac{1}{2},-\frac{1}{2},\frac{1}{2};0,0\right) ,~~\  & \alpha
_{2}=\left( 1,1,0,0,0,0,0,0;0,0\right) , \\ 
\alpha _{3}=\left( -1,1,0,0,0,0,0,0;0,0\right) , & \alpha _{4}=\left(
0,-1,1,0,0,0,0,0;0,0\right) , \\ 
\alpha _{5}=\left( 0,0,-1,1,0,0,0,0;0,0\right) , & \alpha _{6}=\left(
0,0,0,-1,1,0,0,0;0,0\right) , \\ 
\alpha _{7}=\left( 0,0,0,0,-1,1,0,0;0,0\right) , & \alpha _{8}=\left(
0,0,0,0,0,-1,1,0;0,0\right) , \\ 
\alpha _{0}=\left( 0,0,0,0,0,0,-1,-1;1,0\right) , & \alpha _{-1}=\left(
0,0,0,0,0,0,0,0;-1,1\right) , \\ 
\alpha _{-2}=\left( 0,0,0,0,0,0,0,0;0,-1\right) . & 
\end{array}
\label{11roots}
\end{equation}%
We have constructed the analogue of the affine root as $\alpha
_{-2}=-\sum\nolimits_{j=-1}^{8}n_{j}\alpha _{j}$ with Kac labels $%
n=(2,3,4,6,5,4,3,2,1,1,1)$. Using the Lorentzian inner product we compute
for the extended part $\alpha _{-2}^{2}=0$, $\alpha _{-1}^{2}=2$, $\alpha
_{-2}\cdot \alpha _{-1}=-1$, $\alpha _{-1}\cdot \alpha _{0}=-1$, $\alpha
_{-2}\cdot \alpha _{0}=1$. The Dynkin diagram drawn with the standard rules
augmented with the set of rules as stated at the end of the previous
subsection is therefore:

\setlength{\unitlength}{0.58cm} 
\begin{picture}(14.00,3.5)(-1.0,4)
\thicklines
\put(3.5,5.0){\large{$\left(\mathbf{\mathring{E_8}}\right)_{-2}:$}}
\qbezier[18](14.3,5.4)(15.2,6.5)(16.2,5.4)
\put(7.0,5){\Large{$\bullet$}}
\put(7.4,5.2){\line(1,0){0.9}}
\put(6.8,4.5){{$\small{\alpha_{1}}$}}
\put(8.0,5){\Large{$\bullet$}}
\put(8.4,5.2){\line(1,0){0.9}}
\put(8.0,4.5){{$\small{\alpha_{3}}$}}
\put(9.0,5){\Large{$\bullet$}}
\put(9.4,5.2){\line(1,0){0.9}}
\put(9.0,4.5){{$\small{\alpha_{4}}$}}
\put(9.0,6){\Large{$\bullet$}}
\put(9.2,5.2){\line(0,1){0.9}}
\put(9.5,6.2){{$\small{\alpha_{2}}$}}
\put(10.0,5){\Large{$\bullet$}}
\put(10.4,5.2){\line(1,0){0.9}}
\put(10.0,4.5){{$\small{\alpha_{5}}$}}
\put(11.0,5){\Large{$\bullet$}}
\put(11.4,5.2){\line(1,0){0.9}}
\put(11.0,4.5){{$\small{\alpha_{6}}$}}
\put(12.0,5){\Large{$\bullet$}}
\put(12.4,5.2){\line(1,0){0.9}}
\put(12.0,4.5){{$\small{\alpha_{7}}$}}
\put(13.0,5){\Large{$\bullet$}}
\put(13.2,5.2){\line(1,0){0.9}}
\put(12.8,4.5){{$\alpha_{8}$}}
\put(14.0,5){\Large{$\bullet$}}
\put(14.4,5.2){\line(1,0){0.9}}
\put(13.8,4.5){{$\small{\alpha_{0}}$}}
\put(15.0,5){\Large{$\bullet$}}
\put(15.2,5.2){\line(1,0){0.9}}
\put(14.8,4.5){{$\small{\alpha_{-1}}$}}
\put(16.0,5){\Large{$\circ$}}
\put(16.0,4.5){{$\small{\alpha_{-2}}$}}
\end{picture}

The conformal part of the theory is the $\left( \mathbf{E}_{8}\right) _{-1}$%
-theory, aka $\mathbf{E}_{10}$, whose Cartan matrix has exactly one negative
eigenvalue with all other eigenvalues being positive. The Cartan matrix of $(%
\mathbf{\mathring{E}}_{8}\mathbf{)}_{-2}$ has a zero eigenvalue, one
negative eigenvalue with the remaining ones being positive. The mass squared
matrix (\ref{M2}) for the $(\mathbf{\mathring{E}}_{8}\mathbf{)}_{-2}$-theory
is computed to%
\begin{equation}
M^{2}=\frac{1}{2}m^{2}\left( 
\begin{array}{cccccccccc}
15 & -3 & -1 & -1 & -1 & -1 & -1 & 1 & 0 & 0 \\ 
-3 & 27 & -11 & 1 & 1 & 1 & 1 & -1 & 0 & 0 \\ 
-1 & -11 & 23 & -9 & 1 & 1 & 1 & -1 & 0 & 0 \\ 
-1 & 1 & -9 & 19 & -7 & 1 & 1 & -1 & 0 & 0 \\ 
-1 & 1 & 1 & -7 & 15 & -5 & 1 & -1 & 0 & 0 \\ 
-1 & 1 & 1 & 1 & -5 & 11 & -3 & -1 & 0 & 0 \\ 
-1 & 1 & 1 & 1 & 1 & -3 & 7 & 1 & 0 & 2 \\ 
1 & -1 & -1 & -1 & -1 & -1 & 1 & 3 & 0 & 2 \\ 
0 & 0 & 0 & 0 & 0 & 0 & 0 & 0 & 4 & -2 \\ 
0 & 0 & 0 & 0 & 0 & 0 & 2 & 2 & -2 & 4%
\end{array}%
\right) .
\end{equation}%
The ten eigenvalues $%
(19.4794,12.8905,8.8224,7.4524,5.1100,3.7371,3.0181,2.1237$, $1.1227,$ $%
0.2437)$ of $M^{2}$ are all positive, thus leading to a physically
well-defined classical mass spectrum. We may set here $m=1$, as only mass
ratios will be relevant. Similarly we compute the masses for the other
members of the $(\mathbf{\mathring{E}}_{8}\mathbf{)}_{-2n}$-series, which
all posses well defined spectra. We present our results for the first
members of the series in figure \ref{Fig1}.

\FIGURE{ \epsfig{file=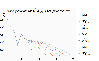, width=12.cm} 
\caption{Mass ratios for the $r+2n$ particles in
the $\left( \mathbf{\mathring{E}}_{8}\right) _{-2n}$-Toda field theories with almost stable noncrystallographic $H_{4}$ compound.}
       \label{Fig1}}

We observe the interesting feature that when comparing the masses with those
of standard $E_{8}$-affine Toda field theory, four masses are especially
stable and remain almost all identical irrespective of the value of $n$.
These masses can be identified when recalling that folding the $E_{8}$%
-affine Toda field theory \cite{fring2005affine} leads to a grouping of the
eight masses in the $E_{8}$-theory \cite{zamo1989} into as two copies of
four masses attributed to a theory based on the root space of
noncrystallographic type $H_{4}$. One set is obtained from the other by a
multiplication of the golden ration $\phi =(1+\sqrt{5})/2$. Normalizing the $%
E_{8}$- masses so that the largest takes on the value $1$, we have 
\begin{eqnarray}
m_{1} &=&1,\quad ~~~~\ \ \ \ m_{2}=2\sin (4\theta ),\quad \ \ m_{3}=\frac{%
\cos \theta }{\phi \cos (4\theta )},\quad \ \ \ \ \ \ m_{4}=\frac{1}{2\phi
\cos (4\theta )},\quad  \label{h4} \\
m_{5} &=&\phi ^{-1}m_{1},\quad m_{6}=\phi ^{-1}m_{2},\quad ~~~\ \ \
m_{7}=\phi ^{-1}m_{3},\quad ~~~\ \ \ \ \ \ \ m_{8}=\phi ^{-1}m_{4},\quad
\end{eqnarray}%
with $\theta =\pi /30$. We observe in figure \ref{Fig1} that the four $H_{4}$
masses in (\ref{h4}) are almost identical in all $(\mathbf{\mathring{E}}_{8}%
\mathbf{)}_{-2n}$-theories.

However, none of these theories, apart from $(\mathbf{\mathring{E}}_{8}%
\mathbf{)}_{0}$, passes the Painlev\'{e} integrability test. In all other
cases the eigenvalues of the matrix $2D_{\mathbf{\mathring{g}}_{-(2n-1)}}K_{%
\mathbf{\mathring{g}}_{-(2n-1)}}$ are all non integer valued and sometimes
negative. We find that $D_{\mathbf{\mathring{g}}_{1}}\equiv D_{\mathbf{E}%
_{8}}$ is positive definite, as is expected for the semi-simple case. We
confirm in this case the relation (\ref{indk}) as $\limfunc{ind}\left( 2D_{%
\mathbf{E}_{8}}K_{\mathbf{E}_{8}}\right) =\limfunc{ind}\left( K_{\mathbf{E}%
_{8}}\right) =8$. Moreover the eigenvalues factorize into $s_{i}(s_{i}+1)$
with $s_{i}=1$, $7$, $11$, $13$, $17$, $19$, $23$, $29$, corresponding to
the $8$ exponents of $E_{8}$.

In contrast, the matrices $D_{\mathbf{\mathring{g}}_{-(2n-1)}}$ are negative
definite for all values of $n\geq 1$. The $8+2n$ eigenvalues for $2D_{(%
\mathbf{\mathring{E}}_{8}\mathbf{)}_{-(2n-1)}}K_{(\mathbf{\mathring{E}}_{8}%
\mathbf{)}_{-(2n-1)}}$for $n=1,2,\ldots $ separate into $8+n$ negative and $%
n $ positive eigenvalues. The relation (\ref{indk}) is confirmed as 
\begin{equation}
\limfunc{ind}\left( -2D_{\mathbf{\mathring{g}}_{-(2n-1)}}K_{\mathbf{%
\mathring{g}}_{-(2n-1)}}\right) =\limfunc{ind}\left( K_{\mathbf{\mathring{g}}%
_{-(2n-1)}}\right) =8,~\ \ \text{for }n=1,2,\ldots
\end{equation}%
Surprisingly the index of $K$ is preserved for all values of $n$. To make
this plausible we list here the first characteristic polynomials $\det
(K-\lambda \mathbb{I})=0$ for the Cartan matrix $K_{\mathbf{\mathring{g}}%
_{-(2n-1)}}$%
\begin{eqnarray}
\limfunc{ch}\left( K_{\mathbf{E}_{8}}\right) &=&\lambda ^{8}-16\lambda
^{7}+105\lambda ^{6}-364\lambda ^{5}+714\lambda ^{4}-784\lambda
^{3}+440\lambda ^{2}-96\lambda +1, \\
\limfunc{ch}\left( K_{(\mathbf{\mathring{E}}_{8}\mathbf{)}_{-1}}\right)
&=&\lambda ^{10}-20\lambda ^{9}+171\lambda ^{8}-816\lambda ^{7}+2379\lambda
^{6}-4356\lambda ^{5}+4949\lambda ^{4}-3304\lambda ^{3} \\
&&+1140\lambda ^{2}-144\lambda -1,  \notag \\
\limfunc{ch}\left( K_{(\mathbf{\mathring{E}}_{8}\mathbf{)}_{-2}}\right)
&=&\lambda ^{12}-22\lambda ^{11}+208\lambda ^{10}-1100\lambda
^{9}+3531\lambda ^{8}-6892\lambda ^{7}+7356\lambda ^{6} \\
&&-1914\lambda ^{5}-4872\lambda ^{4}+5944\lambda ^{3}-2626\lambda
^{2}+388\lambda +1,  \notag \\
\limfunc{ch}\left( K_{(\mathbf{\mathring{E}}_{8}\mathbf{)}_{-4}}\right)
&=&\lambda ^{14}-24\lambda ^{13}+249\lambda ^{12}-1450\lambda
^{11}+5103\lambda ^{10}-10576\lambda ^{9}+9896\lambda ^{8} \\
&&+7088\lambda ^{7}-31796\lambda ^{6}+37074\lambda ^{5}-17467\lambda
^{4}-520\lambda ^{3}+3050\lambda ^{2}-636\lambda -1.  \notag
\end{eqnarray}%
We observe that in each polynomial of the general form $\sum%
\nolimits_{i=1}^{8+2n}a_{i}\lambda ^{i},$ the sequence of coefficients $%
a_{i} $ has exactly $8+n$ sign changes. Thus according to Descartes' rule of
signs, see e.g. \cite{anderson1998}, we have exactly $8+n$ positive real
eigenvalues confirming the observation above. The factorization of these
eigenvalues into $s_{i}(s_{i}+1)$ leads to the form $s_{i}=1/2+\lambda _{i}$
with $\lambda _{i}\in \mathbb{R}$ and $s_{i}=\kappa _{i}$ with $\kappa
_{i}\in \mathbb{R}$, for the negative and positive eigenvalues, respectively.

We depict the eigenvalue spectra for some $\mathcal{L}_{_{(\mathbf{\mathring{%
E}}_{8}\mathbf{)}_{-2n}}}$-extended Lorentzian Toda field theory in figure %
\ref{Fig2}.

\FIGURE{ \epsfig{file=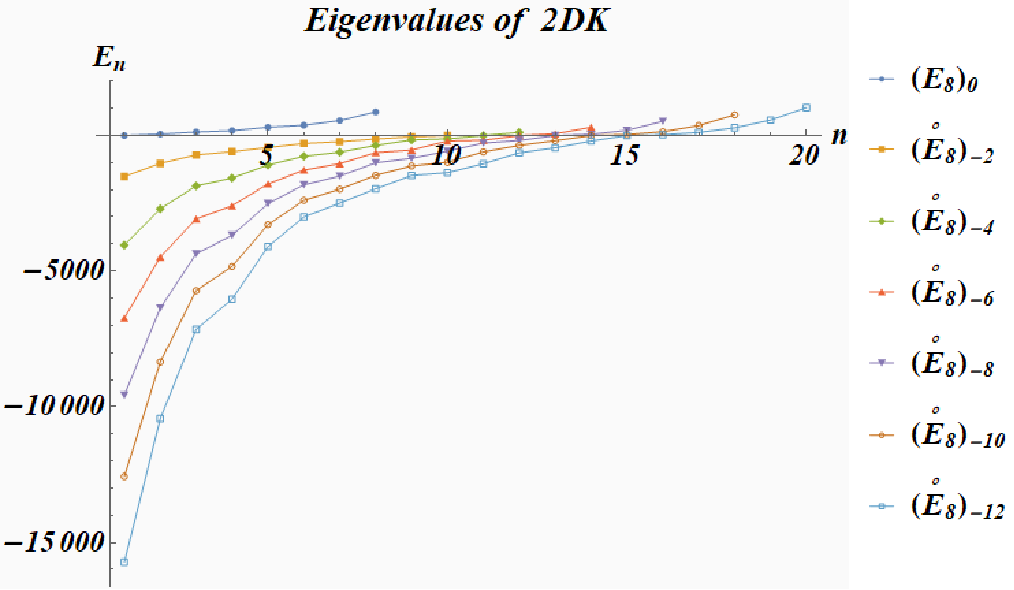, width=12.cm} 
\caption{Eigenvalue
spectra for the Painlev\'{e} matrix $2D_{\mathbf{\mathring{g}}_{-(2n-1)}}K_{\mathbf{\mathring{g}}_{-(2n-1)}}$.}
       \label{Fig2}}

As most of the eigenvalues are negative or non integer valued, the $\mathcal{%
L}_{_{(\mathbf{\mathring{E}}_{8}\mathbf{)}_{-2n}}}$-extended Lorentzian Toda
field theory fail the Painlev\'{e} test and are therefore not integrable.

\subsection{$\mathcal{L}_{(\mathbf{\mathring{g}}_{1}\mathbf{)}_{-2n}\diamond
(\mathbf{\mathring{g}}_{2}\mathbf{)}_{-2m}}$-extended Lorentzian Toda field
theory}

This construction is based on a generalization of what is referred to in 
\cite{gaberdiel2002class} as the symmetric fusion of two finite semisimple
Lie algebras $g_{1}$ and $g_{2}$ by means of some Lorentzian roots in $\Pi
^{1,1}$. Here we consider a root lattice of the form 
\begin{equation}
\Lambda _{(\mathbf{\mathring{g}}_{1}\mathbf{)}_{1-2n}\diamond (\mathbf{%
\mathring{g}}_{2}\mathbf{)}_{1-2m}}=\Lambda _{(\mathbf{\mathring{g}}_{1}%
\mathbf{)}_{1-2n}}\oplus \Pi ^{1,1}\oplus \Lambda _{(\mathbf{\mathring{g}}%
_{2}\mathbf{)}_{1-2m}}.
\end{equation}%
It is comprised of the $r_{1}+2n$ roots $\alpha _{i}$ with $i=1-2n,\ldots
,r_{1}$ of $(\mathbf{\mathring{g}}_{1}\mathbf{)}_{1-2n}$, the $r_{2}+2m$
roots $\beta _{i}$ with $i=1-2m,\ldots ,r_{2}$ of $(\mathbf{\mathring{g}}_{2}%
\mathbf{)}_{-2m}$ and two modified roots 
\begin{equation*}
\alpha _{-2n}=k_{n+1}-\sum\nolimits_{j=1-2n}^{r_{1}}n_{j}\alpha _{j},\qquad
\beta _{-2m}=\bar{k}_{n+1}-\sum\nolimits_{j=-1-2m}^{r_{1}}n_{j}\beta _{j}
\end{equation*}%
with\ $k_{n+1},\bar{k}_{n+1}\in \Pi ^{1,1}$. The Lorentzian roots used in
the construction of the $\alpha $ and $\beta $ roots are unrelated with
mutual inner products equal to zero. They are labeled by $k_{i},\bar{k}_{i}$%
, $i=1,\ldots ,n$ and $\ell _{i},\bar{\ell}_{i}$, $i=1,\ldots ,m$,
respectively. For $n=m=0$ this construction coincides with the one in \cite%
{gaberdiel2002class} apart from a change of sign in the definition of $\beta
_{0}$ where we added $\bar{k}$ instead of $-\bar{k}$ as in \cite%
{gaberdiel2002class}. We explain the reason for our preferred choice below.
The massive version is then constructed by adding a root $\gamma =-(k_{n+1}+%
\bar{k}_{n+1})$. Using the rules as stated above, the part of the Dynkin
diagram where the $(\mathbf{\mathring{g}}_{1}\mathbf{)}_{-2n}$ and $(\mathbf{%
\mathring{g}}_{2}\mathbf{)}_{-2m}$ for $n\geq 1$, $m\geq 1$ are joined is:

\setlength{\unitlength}{0.58cm} 
\begin{picture}(14.00,3.5)(1.5,3.8)
\thicklines

\put(0.5,5.0){\Large{$(\mathbf{\mathring{g}}_{1}\mathbf{)}_{-2n} \diamond (\mathbf{%
\mathring{g}}_{2}\mathbf{)}_{-2m}:$}}
\qbezier[18](11.3,5.4)(12.2,6.5)(13.2,5.4)
\qbezier[18](13.3,5.4)(14.2,6.5)(15.2,5.4)
\qbezier[18](17.3,5.4)(18.2,6.5)(19.2,5.4)
\qbezier[18](19.3,5.4)(20.2,6.5)(21.2,5.4)
\put(9.2,5.2){\Large{$\ldots$}}
\put(10.4,5.2){\line(1,0){0.7}}
\put(11.0,5){\Large{$\circ$}}
\put(11.4,5.2){\line(1,0){0.9}}
\put(10.6,4.5){{$\small{\alpha_{4-2n}}$}}
\put(12.0,5){\Large{$\bullet$}}
\put(12.2,5.2){\line(1,0){0.9}}
\put(13.0,5){\Large{$\circ$}}
\put(13.4,5.2){\line(1,0){0.9}}
\put(12.6,4.5){{$\small{\alpha_{2-2n}}$}}
\put(14.0,5){\Large{$\bullet$}}
\put(14.2,5.2){\line(1,0){0.8}}
\put(15.0,5){\Large{$\circ$}}
\put(14.6,4.5){{$\small{\alpha_{-2n}}$}}
\put(17.0,5){\Large{$\circ$}}
\put(16.6,4.5){{$\small{\beta_{-2m}}$}}
\put(17.4,5.2){\line(1,0){0.9}}
\qbezier[18](15.4,5.2)(16.2,5.2)(17.0,5.2)
\put(16.0,6){\Large{$\bullet$}}
\put(15.3,5.3){\line(1,1){.9}}
\put(17.1,5.3){\line(-1,1){.9}}
\put(16.5,6.3){{$\gamma$}}
\put(18.0,5){\Large{$\bullet$}}
\put(18.2,5.2){\line(1,0){0.9}}
\put(19.0,5){\Large{$\circ$}}
\put(18.6,4.5){{$\small{\beta_{2-2m}}$}}
\put(19.4,5.2){\line(1,0){0.9}}
\put(20.0,5){\Large{$\bullet$}}
\put(20.2,5.2){\line(1,0){0.9}}
\put(21.0,5){\Large{$\circ$}}
\put(20.6,4.5){{$\small{\beta_{4-2m}}$}}
\put(21.4,5.2){\line(1,0){0.7}}
\put(22.4,5.2){\Large{$\ldots$}}

\end{picture}

The corresponding Cartan matrix is simply linking up the two affine Cartan
matrices $K_{(\mathbf{\mathring{g}}_{1}\mathbf{)}_{-2n}}$ and $K_{(\mathbf{%
\mathring{g}}_{2}\mathbf{)}_{-2m}}$ as 
\begin{equation}
K_{(\mathbf{\mathring{g}}_{1}\mathbf{)}_{-2n}\diamond (\mathbf{\mathring{g}}%
_{2}\mathbf{)}_{-2m}}=\left( 
\begin{array}{ccccccccc}
&  &  & q_{1} &  & 0 & 0 & \cdots & 0 \\ 
& K_{(\mathbf{\mathring{g}}_{1}\mathbf{)}_{1-2n}} &  & \vdots &  & \vdots & 
\vdots &  & \vdots \\ 
&  &  & q_{r_{1}} &  & 0 & 0 &  & \vdots \\ 
q_{1} & \cdots & q_{r_{1}+2n} & 0 & -1 & 1 & 0 & \cdots & 0 \\ 
&  &  & -1 & 2 & -1 &  &  &  \\ 
0 & \cdots & 0 & 1 & -1 & 0 & p_{r_{2}+2m} & \cdots & p_{1} \\ 
\vdots &  & \vdots & 0 &  & p_{r_{2}} &  &  &  \\ 
\vdots &  & \vdots & \vdots &  & \vdots &  & K_{(\mathbf{\mathring{g}}_{2}%
\mathbf{)}_{1-2m}} &  \\ 
0 & \cdots & 0 & 0 &  & p_{1} &  &  & 
\end{array}%
\right) ,
\end{equation}%
where $q_{s}:=\alpha _{-2n}\cdot \alpha _{s}$, $s=1,\ldots ,r_{1}+2n$ \ and $%
p_{s}:=\beta _{-2m}\cdot \beta _{s}$, $s=1,\ldots ,r_{2}+2m$.

We present now some examples for Lorentzian Toda field theories build from
concrete algebras of this type of construction.

\subsubsection{$(\mathbf{\mathring{E}}_{8})_{-2n}\diamond (\mathbf{\mathring{%
E}}_{8})_{-2n}$-Lorentzian Toda field theories}

We start with $(\mathbf{\mathring{E}}_{8})_{0}\diamond (\mathbf{\mathring{E}}%
_{8})_{0}\equiv (\mathbf{E}_{8})_{0}\diamond (\mathbf{E}_{8})_{0}$ and take
the same representation for the eight simple roots $\alpha _{i}$, $%
i=1,\ldots 8$ as defined in (\ref{11roots}), but we enlarge the
representation space from $10$ to $18$ dimensions by adding $8$ zero
entries. The modified affine root $\alpha
_{0}=k-\sum\nolimits_{j=1}^{8}n_{j}\alpha _{j}$ takes on the same form as in
(\ref{11roots}). Next we construct the roots for the second set of simple
roots as $\beta _{i}^{j+10}=\alpha _{i}^{j}$, $i,j=1,\ldots 8$, and with all
remaining entries $0$. The second modified affine root is constructed as $%
\beta _{0}=\bar{k}-\sum\nolimits_{j=1}^{8}n_{j}\beta _{j}$. The additional
root $\gamma =-(k+\bar{k})$ has therefore nonvanishing entries $\gamma
^{9}=-\gamma ^{10}=-1$. The Dynkin diagram becomes in this case

\setlength{\unitlength}{0.58cm} 
\begin{picture}(14.00,3.5)(1.5,3.8)
\thicklines
\put(0.5,5.0){\Large{$(\mathbf{E}_8)_0 \diamond (\mathbf{E}_8)_0 $: }}
\put(7.0,5){\Large{$\bullet$}}
\put(7.4,5.2){\line(1,0){0.9}}
\put(6.8,4.5){{$\small{\alpha_{1}}$}}
\put(8.0,5){\Large{$\bullet$}}
\put(8.4,5.2){\line(1,0){0.9}}
\put(8.0,4.5){{$\small{\alpha_{3}}$}}
\put(9.0,5){\Large{$\bullet$}}
\put(9.4,5.2){\line(1,0){0.9}}
\put(9.0,4.5){{$\small{\alpha_{4}}$}}
\put(9.0,6){\Large{$\bullet$}}
\put(9.2,5.2){\line(0,1){0.9}}
\put(9.5,6.2){{$\small{\alpha_{2}}$}}
\put(10.0,5){\Large{$\bullet$}}
\put(10.4,5.2){\line(1,0){0.9}}
\put(10.0,4.5){{$\small{\alpha_{5}}$}}
\put(11.0,5){\Large{$\bullet$}}
\put(11.4,5.2){\line(1,0){0.9}}
\put(11.0,4.5){{$\small{\alpha_{6}}$}}
\put(12.0,5){\Large{$\bullet$}}
\put(12.4,5.2){\line(1,0){0.9}}
\put(12.0,4.5){{$\small{\alpha_{7}}$}}
\put(13.0,5){\Large{$\bullet$}}
\put(13.2,5.2){\line(1,0){0.9}}
\put(12.8,4.5){{$\alpha_{8}$}}
\put(14.0,5){\Large{$\bullet$}}
\put(13.8,4.5){{$\small{\alpha_{0}}$}}
\put(15.0,6){\Large{$\bullet$}}
\put(15.5,6.2){{$\small{\gamma}$}}
\put(14.2,5.2){\line(1,1){.9}}
\put(16.2,5.2){\line(-1,1){.9}}
\qbezier[18](14,5.2)(15.2,5.2)(16,5.2)
\put(16.0,5){\Large{$\bullet$}}
\put(16.0,4.5){{$\small{\beta_{0}}$}}
\put(16.2,5.2){\line(1,0){0.9}}
\put(17.0,5){\Large{$\bullet$}}
\put(17.0,4.5){{$\small{\beta_{8}}$}}
\put(17.2,5.2){\line(1,0){0.9}}
\put(18.0,5){\Large{$\bullet$}}
\put(18.0,4.5){{$\small{\beta_{7}}$}}
\put(18.2,5.2){\line(1,0){0.9}}
\put(19.0,5){\Large{$\bullet$}}
\put(19.0,4.5){{$\small{\beta_{6}}$}}
\put(19.2,5.2){\line(1,0){0.9}}
\put(20.0,5){\Large{$\bullet$}}
\put(20.0,4.5){{$\small{\beta_{5}}$}}
\put(20.2,5.2){\line(1,0){0.9}}
\put(21.0,5){\Large{$\bullet$}}
\put(21.0,4.5){{$\small{\beta_{4}}$}}
\put(21.2,5.2){\line(0,1){0.9}}
\put(21.0,6){\Large{$\bullet$}}
\put(21.5,6.2){{$\small{\beta_{2}}$}}
\put(21.2,5.2){\line(1,0){0.9}}
\put(22.0,5){\Large{$\bullet$}}
\put(22.0,4.5){{$\small{\beta_{3}}$}}
\put(22.2,5.2){\line(1,0){0.9}}
\put(23.0,5){\Large{$\bullet$}}
\put(23.0,4.5){{$\small{\beta_{1}}$}}

\end{picture}

Similarly we construct the Cartan matrix for the other members of the $(%
\mathbf{\mathring{E}}_{8})_{-2n}\diamond (\mathbf{\mathring{E}}_{8})_{-2n}$%
-series.

With a well defined root system and vanishing linear term we can compute the
mass squared matrix as defined in (\ref{M2}). Once more we find that all
eigenvalues of the mass squared matrix are positive. Taking the normalized
square root of these eigenvalues, we depict the classical mass spectra for
the first seven members of the $(\mathbf{\mathring{E}}_{8})_{-2n}\diamond (%
\mathbf{\mathring{E}}_{8})_{-2n}$-series in figure \ref{Fig3}.

\FIGURE{ \epsfig{file=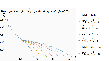, width=12.cm} 
\caption{Mass ratios for the $18+2n$ particles in
the $(\mathbf{\mathring{E}}_{8})_{-2n}\diamond (\mathbf{\mathring{E}}_{8})_{-2n}$-Lorentzian Toda field theory.}
       \label{Fig3}}

\FIGURE{ \epsfig{file=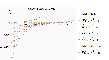, width=12.cm} 
\caption{Real part of the eigenvalue spectra for the $2DK$-matrix for the $(\mathbf{\mathring{E}}_{8})_{-2n}\diamond (\mathbf{\mathring{E}}_{8})_{-2n}$-Lorentzian Toda field theory.}
       \label{Fig4}}

We note that all mass spectra in figure \ref{Fig3} are nondegenerate. Even
though it may appear from the figure that some of the heaviest particles
have the same mass, there is in fact always at least a very small difference
not visible on the scale used in the figure. For the lighter particles in
the spectrum the difference becomes more apparent. Splitting the particles
into sets belonging to the left and right set of roots, $\alpha $ and $\beta 
$, respectively, and comparing with the mass spectrum of the affine $(%
\mathbf{E}_{8})_{0}$-theory, we observe that the mass spectrum of five
heaviest particles is almost identical to the masses in the left and right
set of roots.

Next we consider the eigenvalues of the Painlev\'{e} matrix. First we notice
that the diagonal matrix $D_{(\mathbf{E}_{8})_{0}\diamond (\mathbf{E}%
_{8})_{0}}$ is positive definite and that the relation (\ref{indk}) holds
with $\limfunc{ind}\left( K_{(\mathbf{E}_{8})_{0}\diamond (\mathbf{E}%
_{8})_{0}}\right) =16$. It is these eigenvalue spectrum that motivates the
choice for the sign in front of the Lorentzian roots in the definition of $%
\beta _{0}$. Choosing $-\bar{k}$ instead of $\bar{k}$ will not affect the
mass spectrum, but it will reverse the sign in signature of the eigenvalues
of $2DK$. However, this theory does not pass the Painlev\'{e} integrability
test as the eigenvalues of the matrix $2DK$ are all non integer valued.

In contrast, for $(\mathbf{\mathring{E}}_{8})_{-2n}\diamond (\mathbf{%
\mathring{E}}_{8})_{-2n}$ with $n\geq 1$ the $D$-matrix is semi-definite
with the four central diagonal entries $D_{(9+n\pm 1)(9+n\pm 1)}$, $%
D_{(9+n\pm 2)(9+n\pm 2)}$ being positive and the remaining negative.
Defining a reduced $D$-matrix as $\hat{D}$ by setting the positive entries
to zero we find a reduced version of (\ref{indk}) as $\limfunc{ind}(-2\hat{D}%
K)=\limfunc{ind}(K)=16$. None of the theories in this series passes the
Painlev\'{e} integrability test as the eigenvalues of the matrix $2DK$ are
not only all non integer valued or negative, but in addition some of the
eigenvalues occur in complex conjugate pairs. We depict the real eigenvalues
in figure \ref{Fig4}.

We observe that the \textquotedblleft almost degeneracy is roughly preserved
for the six heaviest particles.

\section{Conclusions}

We have introduced various types of construction principles for conformal,
i.e. massless, and massive versions of Toda field theories based on roots
defined on Lorentzian lattices. We carried out a detailed Painlev\'{e}
integrability test, that established that these theories are in general not
integrable. Nonetheless, the theories possess well defined classical mass
spectra and inherit some of the features of their integrable reductions. For
instance, part of the mass spectrum of the $(\mathbf{\mathring{E}}_{8}%
\mathbf{)}_{-2n}$-theories consists of the four masses of the
noncrystallographic $H_{4}$-theory obtained by folding the integrable affine 
$\mathbf{E}_{8}$-theory. Remarkably, these masses are only slightly changed
for all values of $n$, so that we may view this feature as a remnant that
survives the perturbation of the integrable system.

Evidently there are many interesting routes for further investigations left.
We have only presented here some of the examples of algebras we have
investigated. It would be interesting to extract more generic features from
those and develop an algebraically independent formulation and treatment for
them similar to their integrable counterparts. It is clear from the above,
that there are also more options for possible construction principles that
can be explored further. Of course also standard calculations, such as mass
renormalization for these theories or the study of flows between models can
be carried out.

\medskip

\noindent \textbf{Acknowledgments:} SW is supported by a City, University of
London Research Fellowship.


\begin{thebibliography}{99}
\bibitem{toda1975} M.~Toda, \newblock Studies of a non-linear lattice, %
\newblock Physics Reports \textbf{18}(1), 1--123 (1975).

\bibitem{bogoyavlensky} O.~I. Bogoyavlensky, \newblock On perturbations of
the periodic Toda lattice, \newblock Comm. in Math. Phys. \textbf{51}(3),
201--209 (1976).

\bibitem{yoshida1983} H.~Yoshida, \newblock Integrability of generalized
Toda lattice systems and singularities  in the complex t-plane, \newblock in 
\emph{Nonlinear integrable systems-Classical theory and Quantum  theory},
pages 273--289, World Science Publishing Co Singapore, 1983.

\bibitem{fring2005affine} A.~Fring and C.~Korff, \newblock Affine Toda field
theories related to Coxeter groups of  noncrystallographic type, \newblock %
Nucl. Phys. B \textbf{729}(3), 361--386 (2005).

\bibitem{curt1982conf} T.~L. Curtright and C.~B. Thorn, \newblock %
Conformally invariant quantization of the Liouville theory, \newblock Phys.
Rev. Lett. \textbf{48}(19), 1309 (1982).

\bibitem{mansfield1983light} P.~Mansfield, \newblock Light-cone quantisation
of the Liouville and Toda field theories, \newblock Nucl. Phys. B \textbf{222%
}(3), 419--445 (1983).

\bibitem{BCDS} H.~W. Braden, E.~Corrigan, P.~E. Dorey, and R.~Sasaki, %
\newblock Affine Toda field theory and exact S matrices, \newblock Nucl.
Phys. \textbf{B338}, 689--746 (1990).

\bibitem{Mass2} A.~Fring, H.~C. Liao, and D.~Olive, \newblock The mass
spectrum and coupling in affine Toda theories, \newblock Phys. Lett. B 
\textbf{266}, 82--86 (1991).

\bibitem{toda1} A.~Mikhailov, M.~Olshanetsky, and A.~Perelomov, \newblock %
Two-dimensional generalized Toda lattice, \newblock Comm. Math. Phys. 
\textbf{79}, 473--488 (1981).

\bibitem{OP4} M.~A. Olshanetsky and A.~M. Perelomov, \newblock Quantum
integrable systems related to Lie algebras, \newblock Phys. Rept. \textbf{94}%
, 313--404 (1983).

\bibitem{wilson81} G.~Wilson, \newblock The modified Lax and two-dimensional
Toda lattice equations  associated with simple Lie algebras, \newblock %
Ergodic Theory and Dynamical Systems \textbf{1}(3), 361--380 (1981).

\bibitem{Integ} D.~I. Olive and N.~Turok, \newblock Local conserved
densities and zero curvature conditions for Toda  lattice field theories, %
\newblock Nucl. Phys. \textbf{B257}, 277 (1985).

\bibitem{nt} A.~N. Leznov and M.~V. Savelev, 
\newblock {Two-dimensional exactly and completely integrable dynamical systems
  (monopoles, instantons, dual models, relativistic strings, Lund Regge model,
  genalised Toda lattice et.)}, \newblock Commun. Math. Phys. \textbf{89},
59 (1983).

\bibitem{Holl} T.~Hollowood, \newblock Solitons in affine Toda field theory, %
\newblock Nucl. Phys. \textbf{B384}, 523--540 (1992).

\bibitem{aratyn93} H.~Aratyn, C.~P. Constantinidis, L.~A. Ferreira, J.~F.
Gomes, and A.~H.  Zimerman, \newblock Hirota's solitons in the affine and
the conformal affine Toda models, \newblock Nucl. Phys. B \textbf{406},
727--770 (1993).

\bibitem{David} D.~I. Olive, N.~Turok, and J.~W.~R. Underwood, \newblock %
Solitons and the energy momentum tensor for affine Toda theory, \newblock %
Nucl. Phys. \textbf{B401}, 663--697 (1993).

\bibitem{fring1994vertex} A.~Fring, P.~R. Johnson, M.~A.~C. Kneipp, and
D.~I. Olive, \newblock Vertex operators and soliton time delays in affine
Toda field theory, \newblock Nuclear Physics B \textbf{430}, 597--614 (1994).

\bibitem{toda2} A.~Arinshtein, V.~Fateev, and A.~Zamolodchikov, \newblock %
Quantum S-matrix of the (1 + 1)-dimensional Toda chain, \newblock Phys.
Lett. \textbf{B87}, 389--392 (1979).

\bibitem{zamo1989} A.~B. Zamolodchikov, \newblock Integrals of motion and
S-matrix of the (scaled) $T= T_c$ Ising model  with magnetic field, %
\newblock Int. J. of Mod. Phys. A \textbf{4}(16), 4235--4248 (1989).

\bibitem{CM1} P.~Christe and G.~Mussardo, \newblock Integrable Sytems away
from criticality: The Toda field theory and S  matrix of the tricritical
Ising model, \newblock Nucl. Phys. \textbf{B330}, 465 (1990).

\bibitem{CM2} P.~Christe and G.~Mussardo, \newblock Elastic S-matrices in
(1+ 1) dimensions and Toda field theories, \newblock Int. J. Mod. Phys. 
\textbf{A5}, 4581--4628 (1990).

\bibitem{dorey91} P.~Dorey, \newblock Root systems and purely elastic
S-matrices, \newblock Nucl. Phys. B \textbf{358}(3), 654--676 (1991).

\bibitem{FO} A.~Fring and D.~I. Olive, \newblock The Fusing rule and the
scattering matrix of affine Toda theory, \newblock Nucl. Phys. \textbf{B379}%
, 429--447 (1992).

\bibitem{DDV} C.~Destri and H.~J. de~Vega, \newblock New exact results in
affine Toda field theories: Free energy and wave  function renormalizations, %
\newblock Nucl. Phys. \textbf{B358}, 251--294 (1991).

\bibitem{q1} T.~Oota, \newblock q-deformed Coxeter element in non-simply
laced affine Toda field  theories, \newblock Nucl. Phys. \textbf{B504},
738--752 (1997).

\bibitem{q2} A.~Fring, C.~Korff, and B.~J. Schulz, \newblock On the
universal representation of the scattering matrix of affine  Toda field
theory, \newblock Nucl. Phys. \textbf{B567}, 409--453 (2000).

\bibitem{bernard1991} D.~Bernard and A.~LeClair, \newblock Quantum group
symmetries and non-local currents in 2D QFT, \newblock Comm. in Math. Phys. 
\textbf{142}(1), 99--138 (1991).

\bibitem{gebert1996toda} R.~W. Gebert, S.~Mizogushi, and T.~Inami, \newblock %
Toda field theories associated with hyperbolic Kac-Moody algebra  Painleve
properties and W algebras, \newblock Int. J. of Mod Phys. A \textbf{11}(31),
5479--5493 (1996).

\bibitem{carbone2010} L.~Carbone, S.~Chung, L.~Cobbs, R.~McRae, D.~Nandi,
Y.~Naqvi, and D.~Penta, \newblock Classification of hyperbolic Dynkin
diagrams, root lengths and Weyl  group orbits, \newblock J. of Phys. A:
Math. and Theor. \textbf{43}(15), 155209 (2010).

\bibitem{damour200210} T.~Damour, M.~Henneaux, and H.~Nicolai, \newblock E10
and a small tension expansion of M theory, \newblock Phys. Rev. Lett. 
\textbf{89}(22), 221601 (2002).

\bibitem{nicolai2004low} H.~Nicolai and T.~Fischbacher, \newblock Low level
representations for E10 and E11, \newblock Cont. Math \textbf{343}, 191
(2004).

\bibitem{borcherds1988generalized} R.~Borcherds, \newblock Generalized
Kac-Moody algebras, \newblock Journal of Algebra \textbf{115}(2), 501--512
(1988).

\bibitem{gaberdiel2002class} M.~R. Gaberdiel, D.~I. Olive, and P.~C. West, %
\newblock A class of Lorentzian Kac--Moody algebras, \newblock Nuclear
Physics B \textbf{645}, 403--437 (2002).

\bibitem{fring2019n} A.~Fring and S.~Whittington, \newblock n-Extended
Lorentzian Kac--Moody algebras, \newblock Letters in Mathematical Physics ,
1--22 (2020).

\bibitem{delfino1996non} G.~Delfino, G.~Mussardo, and P.~Simonetti, %
\newblock Non-integrable quantum field theories as perturbations of certain 
integrable models, \newblock Nucl. Phys. B \textbf{473}(3), 469--508 (1996).

\bibitem{babelon1990conformal} O.~Babelon and L.~Bonora, \newblock Conformal
affine sl2 Toda field theory, \newblock Phys. Lett. B \textbf{244}(2),
220--226 (1990).

\bibitem{aratyn1991kac} H.~Aratyn, L.~A. Ferreira, J.~F. Gomes, and A.~H.
Zimerman, \newblock Kac-Moody construction of Toda type field theories, %
\newblock Phys. Lett. B \textbf{254}(3-4), 372--380 (1991).

\bibitem{Painor} P.~Painlev{\'{e}}, \newblock M{\'{e}}moire sur les {\'{e}}%
quations diff{\'{e}}rentielles dont  l'int{\'{e}}grale g{\'{e}}n{\'{e}}rale
est uniforme, \newblock Bull. Soc. Math. France \textbf{28}, 201--261 (1900).

\bibitem{ARS} M.~Ablowitz, A.~Ramani, and H.~Segur, \newblock A connection
between nonlinear evolution equations and ordinary  differential equations
of P-type. II, \newblock J. Math. Phys. \textbf{21}, 1006--1015 (1980).

\bibitem{Pain1} J.~Weiss, M.~Tabor, and G.~Carnevale, \newblock The Painlev{%
\'{e}} property for partial differential equations, \newblock J. Math. Phys. 
\textbf{24}, 522--526 (1983).

\bibitem{Joshi1} N.~Joshi and J.~Peterson, \newblock A method of proving the
convergence of the Painleve expansions of  partial differential equations, %
\newblock Nonlinearity \textbf{7}, 595--602 (1994).

\bibitem{Gramma} B.~Grammaticos and A.~Ramani, \newblock Integrability- and
How to detect it, \newblock Lect. Notes Phys. \textbf{638}, 31--94 (2004).

\bibitem{nicolai2001dio} H.~Nicolai and D.~I. Olive, \newblock The principal
SO (1, 2) subalgebra of a hyperbolic Kac--Moody  algebra, \newblock Lett. in
Math. Phys. \textbf{58}(2), 141--152 (2001).

\bibitem{carrell2005} J.~B. Carrell, \newblock Fundamentals of linear
algebra, \newblock The University of British Columbia (2005).

\bibitem{frappat1989structure} L.~Frappat, A.~Sciarrino, and P.~Sorba, %
\newblock Structure of basic Lie superalgebras and of their affine
extensions, \newblock Comm. in Math. Phys. \textbf{121}(3), 457--500 (1989).

\bibitem{balog1990toda} J.~Balog, L.~Feher, L.~O'Raifeartaigh, P.~Forgacs,
and A.~Wipf, \newblock Toda theory and W-algebra from a gauged WZNW point of
view, \newblock Annals of Physics \textbf{203}(1), 76--136 (1990).

\bibitem{anderson1998} B.~Anderson, J.~Jackson, and M.~Sitharam, \newblock %
Descartes' rule of signs revisited, \newblock The American Mathematical
Monthly \textbf{105}(5), 447--451 (1998).
\end{thebibliography}

\end{document}